\documentclass{article}

\usepackage{arxiv}

\usepackage[utf8]{inputenc} % allow utf-8 input
\usepackage[T1]{fontenc}    % use 8-bit T1 fonts
\usepackage{hyperref}       % hyperlinks
\usepackage{url}            % simple URL typesetting
\usepackage{booktabs}       % professional-quality tables
\usepackage{amsfonts}       % blackboard math symbols
\usepackage{nicefrac}       % compact symbols for 1/2, etc.
\usepackage{microtype}      % microtypography
\usepackage{graphicx}
\usepackage{amssymb}
\usepackage{amsmath}
\usepackage{lineno}
\usepackage{hyperref}
\usepackage{listings}
\usepackage{subfigure}
\usepackage{comment}
\usepackage{enumitem}
\usepackage[dvipsnames]{xcolor}

\title{Deep Preconditioners and their application to seismic wavefield processing}

\author{
  Matteo Ravasi \\
  KAUST\\
  Thuwal, Kingdom of Saudi Arabia \\
  \texttt{matteo.ravasi@kaust.edu.sa}
  }
  
\begin{document}

\chead{Seismic processing with Deep Preconditioners}

\maketitle

\begin{abstract}
  Seismic data processing heavily relies on the solution of physics-driven inverse problems. In the presence of unfavourable data acquisition conditions (e.g., regular or irregular coarse sampling of sources and/or receivers), the underlying inverse problem becomes very ill-posed and prior information is required to obtain a satisfactory solution. Sparsity-promoting inversion, coupled with fixed-basis sparsifying transforms, represent the go-to approach for many processing tasks due to its simplicity of implementation and proven successful application in a variety of acquisition scenarios. Leveraging the ability of deep neural networks to find compact representations of complex, multi-dimensional vector spaces, we propose to train an AutoEncoder network to learn a direct mapping between the input seismic data and a representative latent manifold. The trained decoder is subsequently used as a nonlinear preconditioner for the physics-driven inverse problem at hand. Synthetic and field data are presented for a variety of seismic processing tasks and the proposed nonlinear, learned transformations are shown to outperform fixed-basis transforms and convergence faster to the sought solution.
\end{abstract}

\section{Introduction}

Geophysical inverse problems are notoriously ill-posed and ad-hoc regularisation techniques are usually employed to produce solutions that satisfy our available prior knowledge. A typical example in seismic processing is represented by the problem of \textit{seismic data interpolation} where data acquired by sparsely sampled arrays of receivers are reconstructed onto a regular, finely sampled grid of choice. Interpolation methods can be divided into four main categories: spatial Prediction-Error-Filters (PEFs), wave-equation based, rank-reduction, and domain transform. PEFs interpolation methods \cite{Spitz1991, Liu2011} locally represent seismic data as a superposition of a small number of plane waves; such methods reconstruct aliased data by exploiting their non-aliased low-frequency components. Wave-equation based algorithms \cite{Ronen1987, Fomel2003}, on the other hand, fill gaps in the data by means of an implicit migration-demigration scheme. Transform-based algorithms \cite{Trad2002} exploit the fact that seismic data can be represented by a small number of non-zero coefficients in a suitable transformed domain, whilst acquisition gaps introduce noise in such a domain. Reconstructing missing traces therefore becomes a denoising problem in the transformed domain, which can be tackled by means of direct filtering or sparsity-promoting inversion \cite{Hennenfent2008}. In the latter case, the $L_p$ norm ($p \le 1$) of the reconstructed seismic data in the transformed domain is minimized whilst matching the available traces. The frequency-wavenumber (F-K) transform \cite{Abma2006, Schonewille2009}, Radon transform \cite{Kabir1995, Sacchi1995}, and Curvelet-like transforms \cite{Herrmann2008, Fomel2010, Hauser2012} are examples of successful fixed-basis sparsifying transforms. Alternatively, data-driven sparse dictionaries can be learned directly from the dataset at hand \cite{Zhu2017}. Nevertheless, the success of this family of methods is partially hindered by the slow convergence of most sparsity-promoting optimizers and by the fact that weaker events are usually poorly reconstructed. Finally, rank-reduction methods \cite{Trickett2010, Oropeza2011, Kumar2012, Yang2013} lie their foundation on the observation that fully sampled seismic data exhibit a low-rank pattern that is deteriorated when irregular gaps are introduced in the data. This approach is the generalization of the sparsity-based inversion to matrices, in that minimizing the nuclear norm of a matrix is equivalent to minimizing the $L_1$ norm of a vector containing its eigenvalues.

Despite its enormous popularity, interpolation alone is seldom of interest during a seismic processing project. Combining multiple processing steps into a single inversion is in fact likely to reduce the propagation of errors from one step of processing to the next and increase the processing turnaround time. Being based on the solution of an inverse problem, rank-reduction and transform-based algorithms lend nicely to the introduction of additional physical constraints to further mitigate the ill-posed nature of the seismic interpolation problem. For example, by separating the recorded data into its up- and down-going components as part of the interpolation process, an explicit ghost model can be introduced to provide further physical constraints to the sought solution. Such a strategy can be employed when using either  single-sensor data \cite{Grion2017} or multi-sensor data \cite{Ozbek2010}. Similarly, when receiver-side gradients of the recorded wavefield are also available, additional data terms can be easily included in the problem improving the capabilities of the reconstruction process far beyond the Nyquist sampling criterion \cite{Vassallo2010, Ruan2019}. 

The success of deep learning in various scientific disciplines has recently led to the development of a new family of methods for seismic interpolation. \cite{Mandelli2019} recast the data reconstruction problem as an end-to-end learning task and use the popular U-Net architecture to learn a mapping between the sparsely sampled and fully sampled seismic data. Similarly, \cite{Siahkoohi2018} suggest to use conditional adversarial networks by augmenting the learning process with a discriminator network following the Image-to-Image translation framework of \cite{Isola2017}. More recently, \cite{Kuijpers2021} and \cite{Vasconcelos2022} leverage Recurrent Inference Machines (RIMs), a special type of learned iterative solver \cite{Adler2017} that is specifically designed to solve inverse problems in a data-driven fashion, whilst still including prior knowledge about the forward operator. Such an approach has been shown to outperform end-to-end supervised learning methods and better generalize to out-of-distribution data in various applications including MRI reconstruction \cite{Lonning2019} and imaging of gravitational lenses \cite{Morningstar2019}, to cite a few. 

Nevertheless, all of the aforementioned approaches share the same limitation: they require representative pairs of decimated and fully sampled data, which is usually not available in most seismic processing projects. Whilst relying on synthetic data or field data with similar characteristics (e.g., from  nearby survey) may alleviate the arising of generalization issues, the trained network is usually expected to perform sub-optimally at test time when applied to a different dataset. We refer to \cite{Mandelli2019} for an in-depth analysis of the generalization issues of supervised learning approaches in the context of seismic data reconstruction. So-called domain adaptation techniques (e.g., \cite{ Alkhalifah2021, Birnie2022}) may provide a remedy to this problem; however, such generalization issues have also motivated the development of a second wave of deep learning based algorithms that use neural networks in combination with the known physics of the problem to drive the solution of the inverse problem towards physically plausible solutions. Along these lines, \cite{Kong2020} propose to solve the seismic reconstruction problem in an unsupervised manner using an untrained network as a deep prior preconditioner following the Deep Image Prior concept introduced in \cite{Ulyanov2017}. Whilst this approach circumvents the need for any training data, it is currently hindered by very slow convergence and it is shown to be incapable of recovering strongly aliased events. Anti-aliasing, slope-based regularization \cite{Picetti2021} or a POCS-inspired regularization \cite{Park2020} have been further proposed to increase the interpolation capabilities of such deep prior networks.

Following a different line of thought, \cite{Li2020} suggests to train a nonlinear dimensionality reduction model, such as an AutoEncoder network (AE -- {Kramer1991}), in order to identify latent representations of the expected solution manifold. In a subsequent step, the pre-trained network is used as a regularizer in the solution of a physics-driven inverse problem in the medical imaging context. \cite{Obmann2020} have extended this procedure to sparse AEs and modified the regularization term of the inverse problem to penalize solutions that do not belong to the manifold. Whilst this approach closely resembles classical sparsity promoting inversion schemes with over-complete linear basis functions and provides some theoretical guarantees, it usually renders a more challenging training and inversion compared to the original method. In concurrent work, \cite{Ravasi2021} proposed to use the decoder of the trained AE as a nonlinear preconditioners, solving the inverse problem directly in the latent space of the network. This naturally enforces solutions to be consistent with the manifold of the training data and does not require finding a good balance between the data misfit term and the regularization term. To remark the fact that this preconditioner is composed of a neural network, this approach will be referred to herein as \textit{Deep Preconditioner}.

In this work, we further develop the approach of \cite{Ravasi2021} and show that by carefully designing the AE network architecture, loss function, and the pre-processing pipeline associated with the training data, strong representations can be obtained that ultimately improve the quality of the downstream inversion task. We showcase various applications, ranging from deghosting to wavefield separation with both regularly and irregularly subsampled data, applying them to synthetic datasets of increasing complexity and a marine field dataset. Moreover, contrary to medical applications where a set of representative models is usually available upfront, we show that for seismic applications a representative latent manifold can be identified using data that are not exactly in the same form of the model we wish to invert for. For example, in the problem of joint deghosting and data reconstruction, the AE is trained on the available ghosted data whilst the learned decoder is used to invert the data for a finely sampled, deghosted model vector.

The paper is organized as follows. In Section 2, we introduce the theory of Deep Preconditioners and the training process of the associated AE network. In Section 3, we first apply the proposed approach to a toy problem of one-dimensional signal reconstruction. Subsequently, the same approach is used to solve the joint deghosting and seismic data reconstruction as well as the joint wavefield separation and interpolation problems. In section 4, we discuss a number of avenues for future research, whilst some conclusive remarks are presented in Section 5.

\section{Theory}
An inverse problem is the process of estimating from a given set of observations the unknown underlying factors that originated them. Due to their ill-posed nature, geophysical inverse problems cannot be solved relying on the observations alone, instead additional prior knowledge is also required; such information can be provided either in the form of regularization or preconditioning. In this section, we first recap some basic concepts of the theory of regularized inverse problems. Second, we introduce nonlinear, learned regularizers and preconditioners and discuss how to interchange them with their more commonly used linear counterparts. Finally, the training process devised to learn strong latent representations from a collection of training data is described.

\subsection{Introduction to Inverse Problems}
In this work, we are concerned with finding a stable solution to a linear inverse problem of the form:
\begin{equation}
\label{eq:forward}
\mathbf{y} = \mathbf{G} \mathbf{x},
\end{equation}
where $\mathbf{y} \in \mathbb{R}^n$ and $\mathbf{x} \in \mathbb{R}^m$ are the data and model vectors, respectively, and $\mathbf{G}: \mathbb{R}^m \rightarrow \mathbb{R}^n$ represents a linear mapping between these two vectors. The nature of such an operator will be further detailed for each of the problems considered in this work. Note that the presented framework can be also extended to nonlinear operators.

When the problem is well-posed, the solution of equation \ref{eq:forward} can be obtained by simply inverting the modelling operator: $\hat{\mathbf{x}} = \mathbf{G}^{-1} \mathbf{y}$. For ill-posed problems, prior information must be added to encourage the inversion process to produce realistic solutions; this can be done by approximating equation \ref{eq:forward} with a neighbouring well-posed problem that accommodates for a stable solution. A common approach to include prior knowledge in the inversion process is represented by regularized least-squares inversion:
\begin{equation}
\label{eq:reginverse}
\hat{\mathbf{x}} = \underset{\mathbf{x}} {\mathrm{argmin}} ||\mathbf{y} - \mathbf{G} \mathbf{x} ||_2^2 + \epsilon_R ||\mathbf{Rx}||_p^p,
\end{equation}
where $\mathbf{R}$ is the regularization operator and $||\textbf{x}||_p = (\sum_i |x_i|^p)^{1/p}$ is the $L_p$-norm. Examples of regularization with $p=2$ are the so-called Tikhonov regularization, where $\mathbf{R}$ is chosen to be the identity operator to enforce the norm of the solution to be small, or the Laplacian operator to favour smoothness in the recovered model. Similarly, Total Variation ($p=1$ and $\mathbf{R}=\nabla$) represents a popular choice in the case where the model is expected to be piece-wise constant. Here, $\nabla$ is the gradient operator that computes first-order derivatives of the input vector over its spatial coordinates.

Alternatively, a constrained inverse problem can be created to ensure that the sought after solutions satisfies certain conditions:
\begin{equation}
\label{eq:constrinverse}
\hat{\mathbf{x}} = \underset{\mathbf{x} \in \mathcal{P}} {\mathrm{argmin}} ||\mathbf{y} - \mathbf{G} \mathbf{x} ||_2^2,
\end{equation}
where $\mathcal{P}$ identifies a subspace of $\mathbb{R}^m$ with given properties (e.g., solutions containing only positive values). A common way to solve this problem is to identify a linear projection (or preconditioner) that enforces the solution to belong to the subspace of interest:
\begin{equation}
\label{eq:precinverse}
\hat{\mathbf{z}} = \underset{\mathbf{z}} {\mathrm{argmin}} ||\mathbf{y} - \mathbf{GP} \mathbf{z} ||_2^2 + \epsilon_P ||\mathbf{z}||_p^p,
\end{equation}
where $\mathbf{P} \in \mathbb{R}^z \rightarrow \mathbb{R}^m$ is the so-called preconditioner and $\mathbf{z} \in \mathbb{R}^z$ is the projected variable. Once the minimization problem is solved, the model is finally obtained as $\hat{\mathbf{x}} = \mathbf{P} \hat{\mathbf{z}}$. Smoothness in the solution can be accomplished in this case by using a smoothing operator as preconditioner together with the $L_2$ norm in the penalizing term of the projected variable. Another popular choice of preconditioner is represented by a transformation that projects the model into a possibly over-complete space ($z \ge m$) where the model can be explained by a small number of non-zero coefficients. This concept lies at the basis of so-called sparsity-promoting inversion where the $L_1$ norm of the projected variable is minimized alongside the data term to enforce sparsity in the obtained solution \cite{Candes2006, Hennenfent2008}. 

\subsection{Deep Regularization and Preconditioning}
In both of the approaches discussed so far, the choice of the regularizer and preconditioner is generally driven by experience and it can take a great deal of human effort and time to find a suitable transform for a specific problem. Whilst dictionary learning can partially overcome this limitation \cite{Zhu2017}, linear bases struggle to accurately approximate complex vector spaces such as those spanned by seismic data. This is however not the case for nonlinear dimensionality reduction techniques that leverage deep neural networks as we will see in the following. 

To begin with, equations \ref{eq:reginverse} and \ref{eq:precinverse} must be adapted to accommodate for such nonlinear transformations. Starting from the regularized inverse problem, equation \ref{eq:reginverse} can be recast as (e.g., \cite{Obmann2020}):
\begin{equation}
\label{eq:aereginverse}
\hat{\mathbf{x}} = \underset{\mathbf{x}} {\mathrm{argmin}} ||\mathbf{y} - \mathbf{G} \mathbf{x} ||_2^2 + \epsilon_R ||F_\theta(\mathbf{x})-\mathbf{x}||_p^p,
\end{equation}

Alternatively, %by leveraging the same neural network architecture and training process, 
the preconditioned inverse problem in equation \ref{eq:precinverse} can be rewritten as \cite{Ravasi2021}:
\begin{equation}
\label{eq:aeprecinverse}
\hat{\mathbf{z}} = \underset{\mathbf{z}} {\mathrm{argmin}} ||\mathbf{y} - \mathbf{G} F'_\theta(\mathbf{z}) ||_2 + \epsilon_Z ||\mathbf{z}||_p,
\end{equation}
where $F_\theta: \mathbb{R}^m \rightarrow \mathbb{R}^m$ and $F'_\theta: \mathbb{R}^z \rightarrow \mathbb{R}^m$ are nonlinear transformations, possibly represented by a neural network with learned parameters $\theta$. Whilst this network can be obtained in different ways, in this work we will consider AE network architectures. In the first case, $F_\theta(\mathbf{x})= D_\theta(E_\phi(\mathbf{x}))$ where $E_\phi$ and $D_\theta$ are the encoder and decoder parts of the network, respectively. The regularization term in equation \ref{eq:aereginverse} does therefore assess the similarity between a given solution of the inverse problem, $\hat{\textbf{x}}$, and the output of the network to which such solution is fed as input. When the solution is not part of the manifold of expected models (identified by the training data), the regularization term will be large and therefore it will drive the inverse process away from such a solution and towards a more representative model vector that minimizes the data misfit term at the same time. In the latter case, the network simply becomes the decoder of the AE architecture (i.e.,  $F'_\theta=D_\theta$). Here, similarly to traditional linear preconditioners, the decoder takes any vector $\textbf{z}$ in the latent space and transforms it into a vector $\textbf{x}$ in the original space. This naturally ensures that the produced vector $\textbf{x}$ belongs to the manifold of expected solutions present in the training data. Note that in both cases, since the functionals in equations \ref{eq:aereginverse} and \ref{eq:aeprecinverse} are nonlinear, nonlinear solvers are required to estimate the latent variable $\textbf{z}$. The second-order, nonlinear L-BFGS solver \cite{Nocedal1980} is used in this work. Moreover, the choice of the starting guess $\textbf{z}_{0}$ represents an important factor in the success of the inversion process as we will discuss in the Numerical examples section.

\subsection{AutoEncoder training}
In both of the above scenarios, the solution of the inverse problem in equations \ref{eq:aereginverse} and \ref{eq:aeprecinverse} is formulated as a two-steps process as depicted in Figure \ref{fig:workflow}: first, an AE network is trained to learn a latent representation from a set of training data that share high-level features with the expected solution of the inverse problem at hand. Subsequently the entire network (in the regularized case) or the decoder part of the network (in the preconditioned case) is used to drive the solution of the inverse problem towards a solution that belongs to the manifold of the training data.

Starting from the training phase, this process is performed by using a training dataset of $n_s$ samples, $X=\{\mathbf{x}_0, \mathbf{x}_1,..., \mathbf{x}_{n_s}\}$, and it is accomplished by minimizing the following cost function:
\begin{equation}
\label{eq:aetraining}
\hat{\theta} = \underset{\phi, \theta}  {\mathrm{argmin}} \frac{1}{n_s} \sum_{j=1}^{n_s}
\mathcal{L}(\mathbf{x}_j, D_\theta(E_\phi(\mathbf{x}_j))) + \epsilon_E ||E_\phi(\mathbf{x}_j)||_p^p,
\end{equation}
where $\hat{\mathcal{L}}$ is the main loss function, and a regularization term acting on the latent space vectors $\mathbf{z}_j=E_\phi(\mathbf{x}_j)$ is also introduced. More specifically, the loss function can be expressed as:
\begin{equation}
\label{eq:lis}
\mathcal{L}(\mathbf{x}_j, D_\theta(E_\phi(\mathbf{x}_j))) =  \sum_{i=1}^{n_l}
\hat{\mathcal{L}}_i(\mathbf{x}_j, D_\theta(E_\phi(\mathbf{x}_j))),
\end{equation}
where $\hat{\mathcal{L}}_i$ represents the i-th loss function, and $n_l$ indicates that multiple losses can be used to form the overall cost function.

The training process of an AE network is usually impacted by the choice of the network architecture, loss function(s), and training data (and their associated pre-processing). In the subsequent subsections, we highlight a number of strategies that have been adopted in our numerical examples to enhance the reconstruction capabilities of the AE network, whilst ultimately producing more expressive latent representations. This will in turn impact the quality of our downstream task (e.g., deghosting and interpolation). As such, the effectiveness of the different training strategies will be evaluated both by computing the Mean Square Error ($MSE=\sqrt{||\textbf{x}-\hat{\textbf{x}}||_2^2} / m$) of the reconstructed samples in the validation dataset, as well as by computing the signal-to-noise ratio ($SNR=10log_{10}(||\textbf{x}||_2^2/||\textbf{x}-\hat{\textbf{x}}||_2^2)$) of the estimated wavefield of the downstream processing task against the ground truth solution.

\subsubsection{Network Architecture}
When dealing with gridded, multi-dimensional signals such as natural images or seismic data, Convolutional Neural Networks (CNNs) represent the most natural choice for the network architecture. Convolutional AEs \cite{Masci2011} are usually composed of a number of convolutional layers followed by downsampling (implemented via average or max pooling) in the encoder (or contracting) path, and similarly by a number of convolutional layers followed by upsampling (implemented via, for example, bilinear interpolation) in the decoder (or expanding) path; moreover, a $1 \times 1$ convolutional layer or a dense layers may be used at the end of the encoder path and at the start of the decoder path to transform the convolutional features into a vector of size $z$ (i.e., the latent code). Here, we use the latter choice of layer. A final convolutional layer is also added to the decoder to restore the number of channels to 1 like in the input data.

\begin{figure*}
  \centering
  \includegraphics[width=0.75\textwidth]{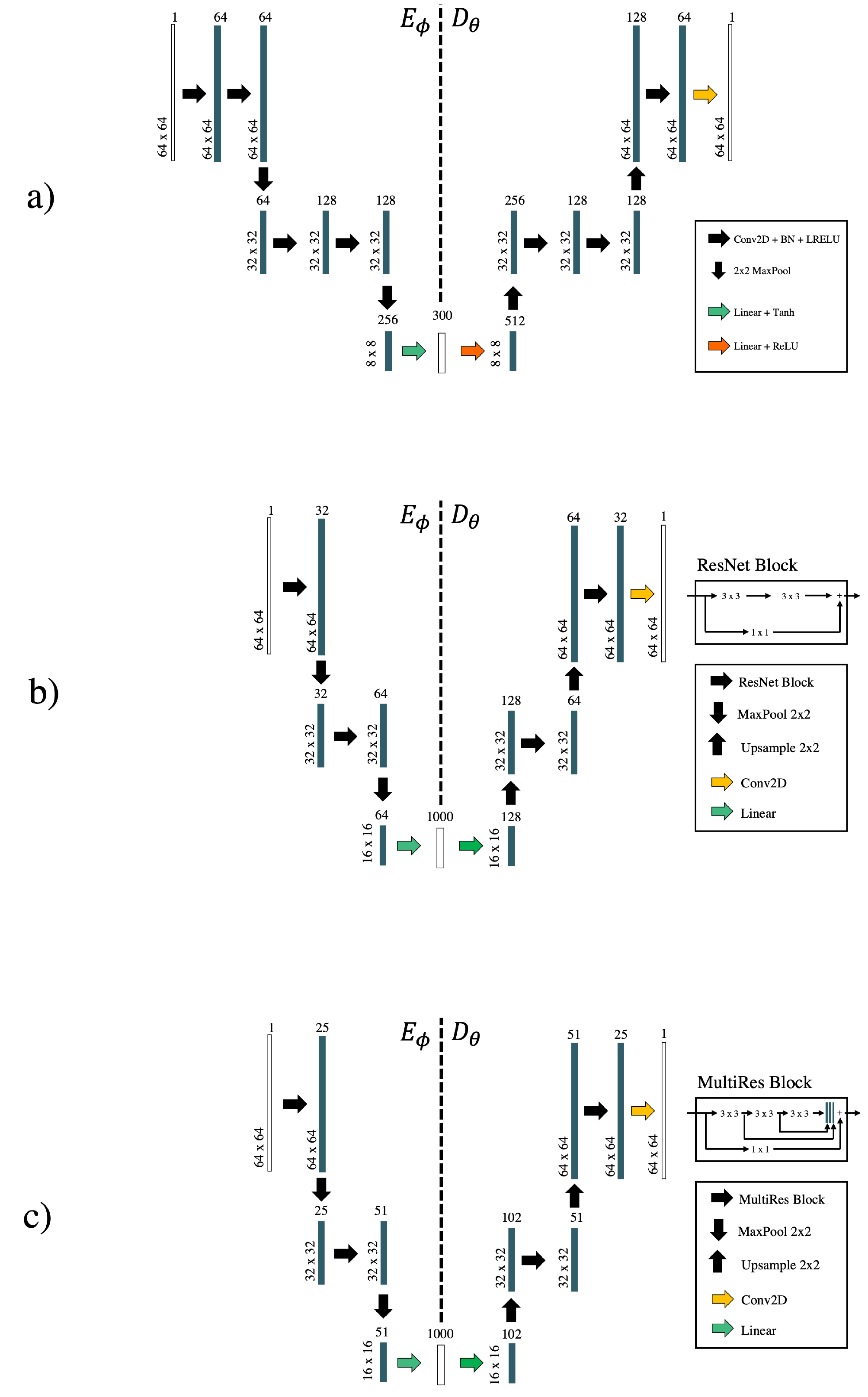}
  \caption{Network architectures. a) Convolutional AE, b) Convolutional AE with ResBlocks, c) Convolutional AE with MultiRes blocks.}
  \label{fig:networks}
\end{figure*}

In this work, a purely convolutional network architecture is chosen as baseline (Figure \ref{fig:networks}a). Each block in the contracting and expanding paths is composed of two convolutional layers followed by batch normalization and a Leaky Rectified Linear Unit (or Leaky ReLU) activation function (with $\alpha=0.2$). A hyperbolic tangent (TanH) activation function is used for the dense layer in the encoding path to ensure boundness of the latent space, whilst a ReLU activation function is used for the dense layer in the decoding path. Max pooling with stride of 2 is used in the encoding path and bilinear interpolation with upsampling factor of 2 is chosen for the decoding path. Two additional network architectures are also considered: the first replaces all the convolutional blocks with residual blocks, or ResBlocks (\cite{He2015} -- Figure \ref{fig:networks}b), whilst the latter uses MultiRes blocks (\cite{Kong2020} -- Figure \ref{fig:networks}c), which have shown promise in the context of deep image prior bases seismic interpolation.

\subsubsection{Loss function}
The choice of the loss function involved in the training process is usually dictated by the expected statistics of the noise present in the training data. More specifically, the MSE loss ($\mathcal{L}_{MSE}(\textbf{x}, \hat{\textbf{x}})= || \textbf{x} - \hat{\textbf{x}}||_2^2$) is chosen in the presence of white, Gaussian noise, whilst the Mean Absolute Error (MAE) loss ($\mathcal{L}_{MAE}(\textbf{x}, \hat{\textbf{x}})=|| \textbf{x} - \hat{\textbf{x}} ||_1$) is preferred in the presence of Laplace noise. However, such losses are local in the sense that they measure the element-wise difference between each value of the predicted and target training samples. Whilst signal fidelity is of great importance in seismic applications, local losses may be unable to provide useful feedback to the network in terms of higher-level characteristics of the signal of interest. In the context of natural images, \cite{Bergmann20158} suggested to use a perceptual loss function based on structural similarity index (SSIM) to captures inter-dependencies between local regions of the predicted and target samples. Along similar lines, \cite{Ovcharenko2021} suggested to combine point-wise losses and trace-wise correlation coefficients as a way to inform the training process about both local and global features of the target data. Examples of trace-wise, scale-independent losses are the Pearson correlation coefficient or the Concordance correlation coefficient (CCC), the latter defined as:
\begin{equation}
\label{eq:ccc}
CCC(\textbf{x}, \hat{\textbf{x}}) = \frac{2\rho_{\textbf{x}, \hat{\textbf{x}}}  \sigma_{\textbf{x}} \sigma_{\hat{\textbf{x}}} } {\sigma_\textbf{x}^2 + \sigma_{\hat{\textbf{x}}}^2 + (\mu_{\textbf{x}} - \mu_{\hat{\textbf{x}}})^2},
\end{equation}
where $\mu_\textbf{x}$ and $\sigma_\textbf{x}$ are the mean and standard deviations of $\textbf{x}$, respectively (and similarly for the vector $\hat{\textbf{x}}$), and $\rho_{\textbf{x}, \hat{\textbf{x}}}$ is the correlation coefficient between the two variables. Note that since CCC tends to zero for uncorrelated signals and one for correlated signals, we define the loss to minimize as, $\mathcal{L}_{CCC}(\textbf{x}, \hat{\textbf{x}}) = 1 - CCC$.

In this work we follow this second strategy and combine the MSE and CCC losses together. Instead of choosing their relative weighting upfront, each loss is equipped with a learned weighting factor $\sigma_i$ and defined as follows:
\begin{equation}
\label{eq:augloss}
\hat{\mathcal{L}}_i = \frac{1}{2 \sigma_i^2} \mathcal{L}_i + log \sigma_i,
\end{equation}
following the multi-task loss function proposed by \cite{Kendall2018}. 
Intuitively $\sigma_i$ quantifies the complexity associated with the i-th task. The network is therefore naturally encouraged to learn the easy task first and tackle the harder task later. The network achieves this by initially increasing the weight of the loss associated to the hard task, which effectively reduces the contribution of the associated gradient into the minimization of the multi-objective functional. The network is however not allowed to completely ignore a task as that would require increasing the associated weight to infinity: this is avoided by the presence of the logarithmic term. Putting all together for our specific problem, the loss function in equation \ref{eq:lis} becomes:
\begin{equation}
\mathcal{L}(\textbf{u}, \hat{\textbf{u}})=
\frac{1}{2\sigma_1^2} \mathcal{L}_{MSE}(\textbf{u}, \hat{\textbf{u}}) + \frac{1}{2\sigma_2^2} \mathcal{L}_{CCC}(\textbf{u}, \hat{\textbf{u}}) + log(\sigma_1 \sigma_2)
\label{eq:multiloss}
\end{equation}
%where $w_i$ represent the weights of the two losses and $\hat{w}_i = 0.5 e^{-w_i}$. Such weights are not fixed a-priori, instead they are learned alongside the network parameters $\theta$. Note that the last term is responsible for avoiding the network to collapse to zero. In practice, during the learning process the network is encourage to first learn to simple task and subsequently tackle the other task. \textbf{[NEED TO READ THE PAPER AND GRAB SOME TEXT THERE...]}

\subsubsection{Pre-processing}
In order to identify robust representations and avoid the network to learn the identity mapping, various regularization strategies have been proposed in the literature for training of AE networks. Denoising AEs \cite{Vincent2008} partially corrupt the input vectors by adding noises to or masking some of their values in a stochastic manner. The target is however kept unchanged. This design is motivated by the fact that humans can easily recognize objects even when they are partially occluded or corrupted because they are able to focus on the key characteristic of such objects. Similarly, an AE can successfully learn robust latent representations only when it is forced to discover and capture high-level relationships in the input data whilst ignoring its missing parts. Since we are mostly interested in reconstructing missing gaps in seismic data as part of the downstream processing task, we follow the second procedure and randomly mask $20 \%$ of the traces from each training sample. This is done differently from epoch to epoch. %Note that the traces removed from one specific training sample changes from epoch to epoch. 
This approach is becoming very popular in the machine learning community in the context of self-supervised learning for both text (e.g., \cite{Devlin2018}) and image (e.g., \cite{He2021}) analysis.

\begin{figure*}
  \centering
  \includegraphics[width=0.99\textwidth]{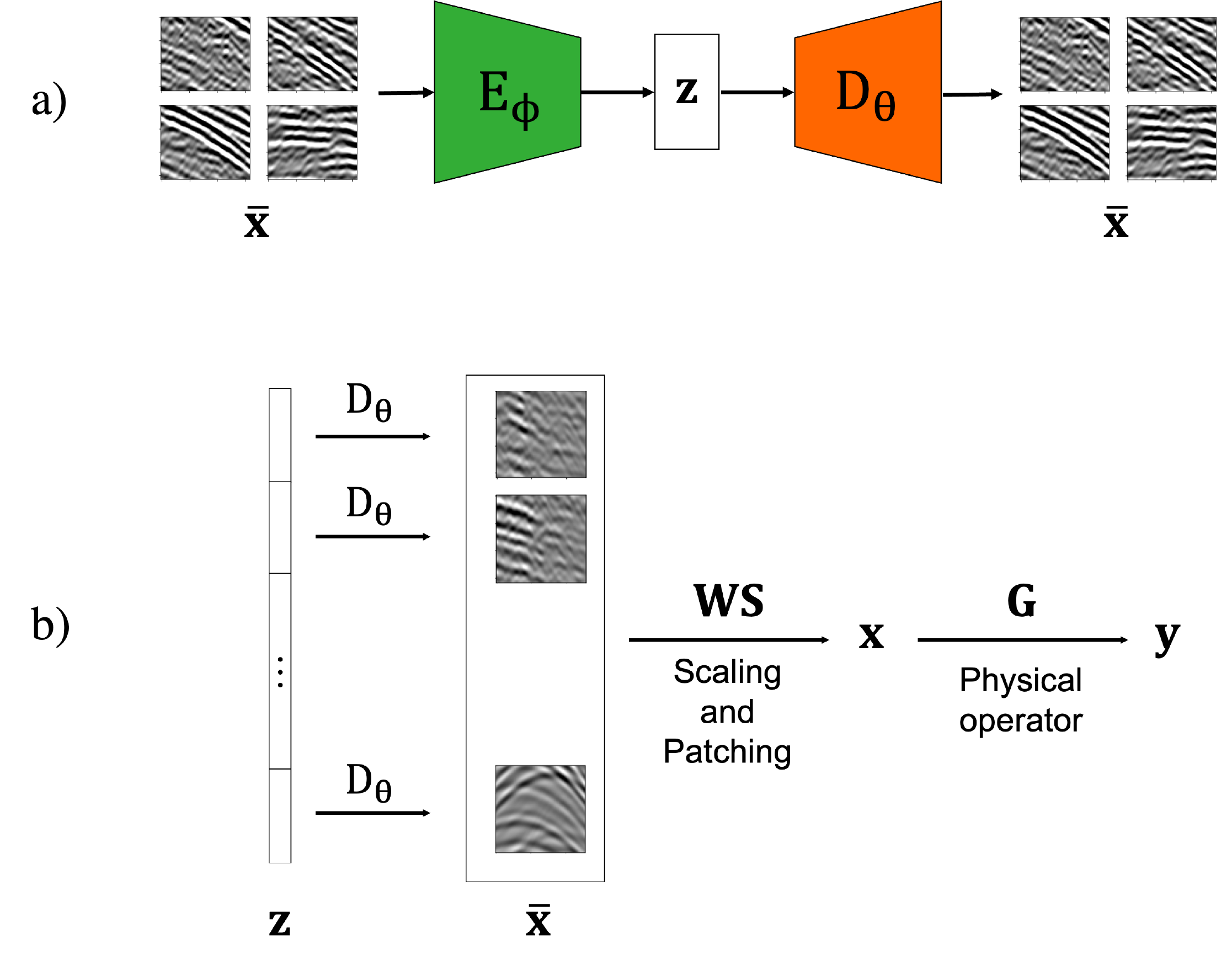}
  \caption{Schematic representation of the proposed two-step approach for the solution of geophysical inverse problems. a) Training phase: an AE network is trained to reproduce patches of seismic data with the aim of learning robust latent representations. b) Inversion phase: the pre-trained decoder is coupled to a physical modelling operator $\textbf{G}$ to solve a geophysical inverse problem of choice. Note that since training has been performed using patches, the latent vector $\textbf{z}$ is composed of a stack of latent codes from different patches of the model vector that we wish to reconstruct. Moreover, since during training all patches have been scaled between $[-1,1]$ an adaptive scaling operator $\textbf{S}$ is applied to the different patches before combining them together via a patching operator $\textbf{W}$.}
  \label{fig:workflow}
\end{figure*}

\section{Numerical examples}

\subsection{Toy example: reconstruction of one-dimensional sinusoidal signals}
The proposed methodology is initially applied to a 1D, sinusoidal signal with the aim of interpolating irregularly spaced gaps in the data. Despite its apparent simplicity, this inverse problem is severely ill-posed and requires prior knowledge of the sought model vector in order to be able to fill the gaps in the recorded signal. This examples is therefore aimed at providing the reader with an intuitive understanding of the value of finding a suitable nonlinear latent representation of the model vector to solve problems in the form of equation  \ref{eq:aeprecinverse}. As a comparison we also solve the reconstruction problem with:
\begin{itemize}
\item a regularizer that penalizes the second-order derivative of the model (i.e., enforces smoothness) using equation \ref{eq:reginverse};
\item a preconditioner based on a linear dimensionality reduction technique using equation \ref{eq:precinverse}. Here, Principal Component Analysis (PCA -- \cite{Hotelling1933}) is chosen as the dimensionality reduction technique.
\end{itemize}

\begin{figure*}
  \centering
  \includegraphics[width=0.99\textwidth]{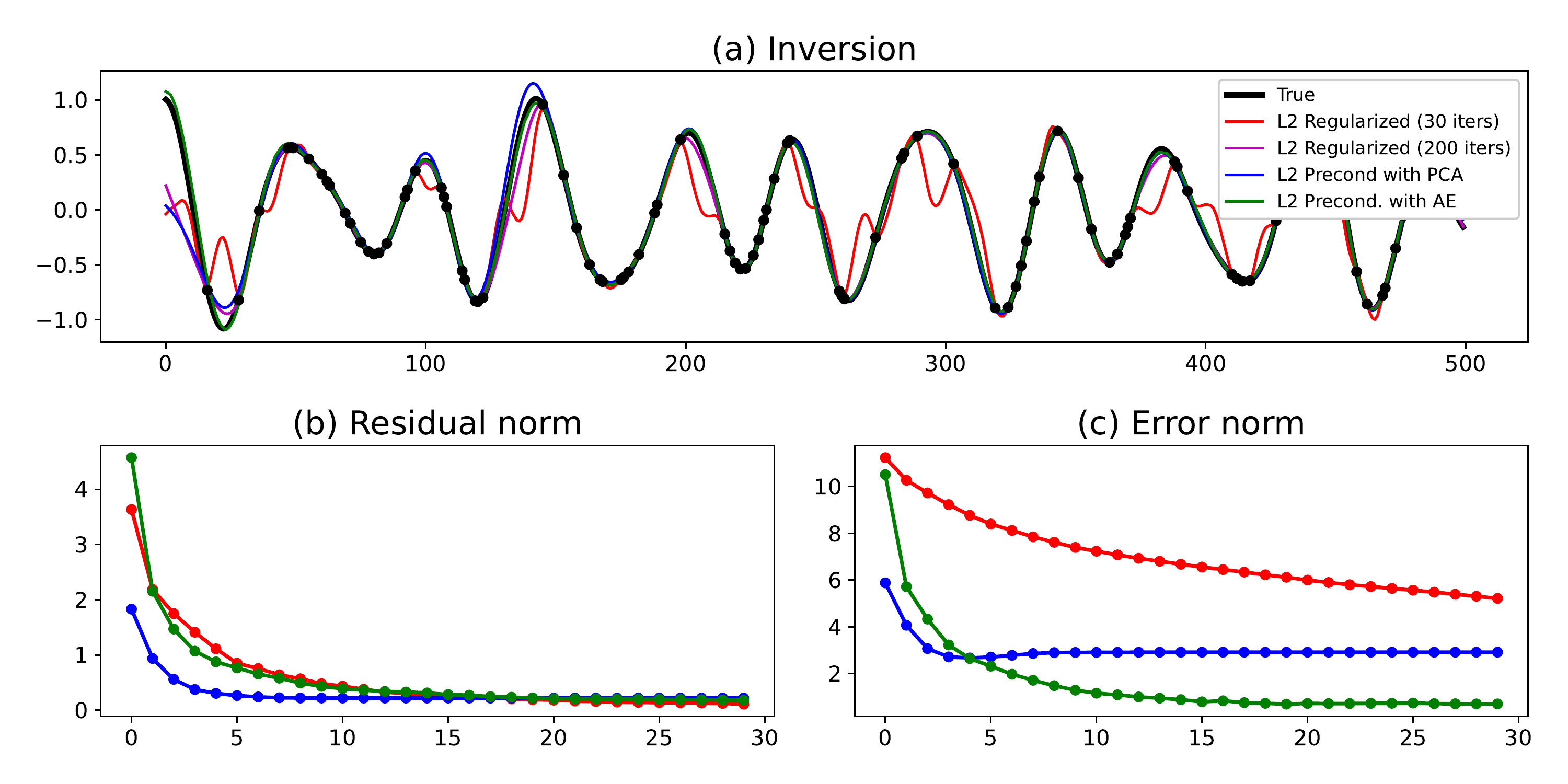}
  \caption{a) Sinusoidal signal reconstruction for the different inversion approaches. b) Residual norm and c) error norm as function of iterations.}
  \label{fig:sinusoid}
\end{figure*}

The forward problem is defined by a restriction operator that extracts values from the finely sampled signal $\textbf{x} \in \mathbb{R}^{500}$ (black line in Figure \ref{fig:sinusoid}a) at irregular locations to form the data vector $\textbf{y} \in \mathbb{R}^{100}$ (black dots in Figure \ref{fig:sinusoid}a). To train both dimensionality reduction techniques, we assume that our signal originates from a parametric family of curves: $x(t)=\sum_{i=1}^N a_i sin(2\pi f_i t + \phi_i)$ where $N$, $a$, $f$ and $\phi$ are sampled from uniform distributions. We sample 30000 curves and split them as follows: 90\% for training and 10\% validation. Based on trial-and-error, the dimensionality of the latent space is chosen to be equal to $k=40$. The encoder and decoder are fully connected neural networks composed of two layers, and both of their hidden layers have size of 80. The Rectified Linear Unit (or ReLU) activation function is used for the hidden layers of both networks, apart from the last dense layer in the encoding path where a hyperbolic tangent (TanH) activation function is chosento ensure boundness of the latent space. Training is performed using the Adam optimizer \cite{Kingma2014} with learning rate $l_r=10^{-3}$, and weight decay regularization $\epsilon_\theta=10^{-5}$, using a single MSE loss in equation \ref{eq:aetraining}. After 15 epochs, the reconstruction error for both the train and validation set is virtually zero. The trained decoder is used to solve equation \ref{eq:aeprecinverse} with 30 iterations of L-BFGS (green line in Figure \ref{fig:sinusoid}a). This solution is compared to that of the regularized problem after 30 (red line in Figure \ref{fig:sinusoid}a) and 200 (magenta line in Figure \ref{fig:sinusoid}a) iterations of LSQR, respectively, and to the solution of the PCA preconditioned problem after 30 iterations of LSQR (blue line in Figure \ref{fig:sinusoid}a). Given the simplicity of the problem, the initial guess $\textbf{z}_{0}$ (or $\textbf{x}_{0}$) is chosen equal to the null vector in all cases. Faster convergence is observed in terms of the residual norm for both preconditioned solutions compared to the regularized ones (Figure \ref{fig:sinusoid}b). More importantly, the error norm of the regularized solution decays very slowly compared to their preconditioned counterparts. 

A major difference is also observed between the PCA and AE error norms: the former plateaus at around 2.5 after a few iterations, whilst the latter goes to zero after about 20 iterations. In other words, the latent representation found by means of PCA is not able to fully capture the signal we wish to recover, whilst that of the AE is more successful at mitigating the ill-posed nature of the inverse problem and drive the nonlinear optimizer to a satisfactory solution.

\subsection{Synthetic seismic datasets}
\begin{figure*}
  \centering
  \includegraphics[width=0.99\textwidth]{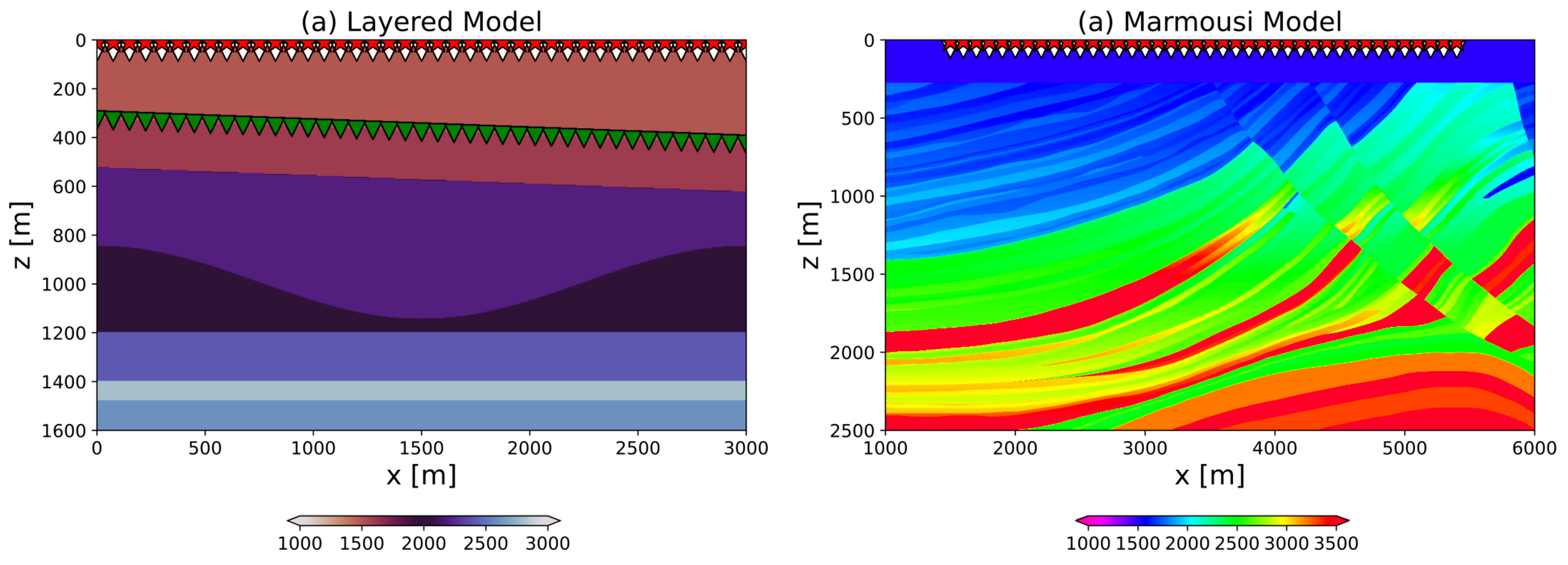}
  \caption{Velocity models. a) Layered model with dipping seabed. Sources (red triangles) and receivers are placed inside the water column (white triangles) and along the seafloor (green triangles). b) Marmousi model with sources and receivers located inside the water column.}
  \label{fig:models}
\end{figure*}

We turn now our attention onto the various seismic datasets that will be used in the following numerical examples. The first synthetic dataset is modelled using a rather simple layered medium (Figure \ref{fig:models}a) using a first-order, staggered-grid, acoustic finite-difference modelling code. The acquisition geometry is composed of 201 sources spaced every 15m at a depth of 10m below the free-surface. Two receiver arrays are placed at a depth of 50m below the free-surface and along a dipping seabed, respectively, both with receiver sampling equal to 15m. The dataset is modelled using a Ricker wavelet with $f_{dom}=15 Hz$ and two subsampled versions of it are created by decimating the receivers as follows: i) irregularly, by a factor of 30\%, or ii) regularly, by keeping one receiver every 4 (25\% available data).
%This dataset will be later used to assess the capabilities of the proposed method to jointly deghost and interpolate seismic data as well as to perform joint multi-component wavefield separation and data reconstruction.
A second synthetic dataset is created using a more realistic geological model obtained by adding a water column of 275m to the Marmousi model (\cite{Brougois1990} -- Figure \ref{fig:models}b]). The acquisition geometry is composed of 199 sources spaced every 20m at a depth of 10m below the free-surface. A receiver array is placed at a depth of 20m below the free-surface with receiver sampling equal to 20m. The dataset is modelled using a Ricker wavelet with $f_{dom}=15 Hz$ and also subsampled as follows: i) irregularly by a factor of 40\%, ii) regularly by keeping one receiver every 3 (33\% available data). Finally, we consider the openly available Mobil AVO viking graben line 12 field dataset\footnote{See https://wiki.seg.org/wiki/Mobil\_AVO\_viking\_graben\_line\_12 for details.}. This dataset has been collected using streamer acquisition system that contains 1001 sources ($dx_S=25m$) and 120 receivers ($dx_R=25m$) with minimum offset equal to $262m$. In our experiment, the dataset is further by randomly selecting 60\% of the available receivers.

\subsection{AutoEncoder training}
In this section, we discuss the training process of the AE network performed on the first synthetic dataset. To begin with, the dataset is sorted into common receiver gathers (CRGs) for all of the available receivers; overlapping patches of size $64 \times 64$ are created (a $50\%$ overlap in both time and space is used in our experiments), and a data augmentation strategy is employed to increase the number of patches that contain events near the direct arrival; this is usually the arrival with largest slope and therefore exhibiting stronger aliasing effects in poorly sampled datasets. Here, we randomly select the center of the patch in a window around the traveltime of the direct arrival and each patch is also flipped horizontally to double the number of patches with similar characteristics. As discussed in more details later in the paper, the main assumption made by our methodology is that we have access to well sampled seismic data in one domain of choice that is representative for the data in the poorly sampled domain. As seismic data are usually acquired in configurations where either the source or the receiver arrays are well sampled, exploiting reciprocity is a well known strategy, not only in this context but also in other traditional or data-driven \cite{PicettiThesis} seismic interpolation methods. 

Each patch is normalized by their absolute maximum value to ensure a maximum dynamic range of $(-1,1)$ for all of the available patches. The entire dataset is composed of 30k patches, which, similarly to the previous example, are randomly split into train (90\%) and validation (10\%) sets. Although the training process is fully unsupervised, the choice of retaining a number of patches for validation is motivated by the fact that we want to assess a variety of training strategies and compare them in terms of their reconstruction capabilities. The training process is performed using the Adam optimizer for 20 epochs, using an initial learning rate equal to $1e^{-4}$ in all experiments and modified during the training process using a scheduler: on-plateau scheduling is selected for the networks with single loss function, whilst the one-cycle scheduling \cite{Smith2017} is chosen for the networks using multiple losses with learned weights. Finally, independent of the network architecture, the latent space vector is chosen as $\textbf{z} \in \mathbb{R}^{300}$, which represents a 13.6x compression factor over the size of the original space, $\textbf{x} \in \mathbb{R}^{64 \times 64}$.

%A properly selected learning rate policy improves the convergence rate and leads the iterative non-linear optimization to a deeper minimum of the objective function. The concept of super-convergence introduced by [167] suggests using the one-cycle strategy, which changes the learning rate for every batch, gradually increasing it from the initial warm-up pace to a large maximum rate, and then decreases it back to an ever lower value than the initial rate. The authors show that the large learning rate serves as an auxiliary regularizer when approached following the one-cycle strategy. We too observe in our experiments that the training reaches a deeper minimum of the objective function when using the one-cycle policy, rather than the reduce-on-the-plateau strategy that quickly overfits our data.

\begin{figure*}
  \centering
  \includegraphics[width=0.99\textwidth]{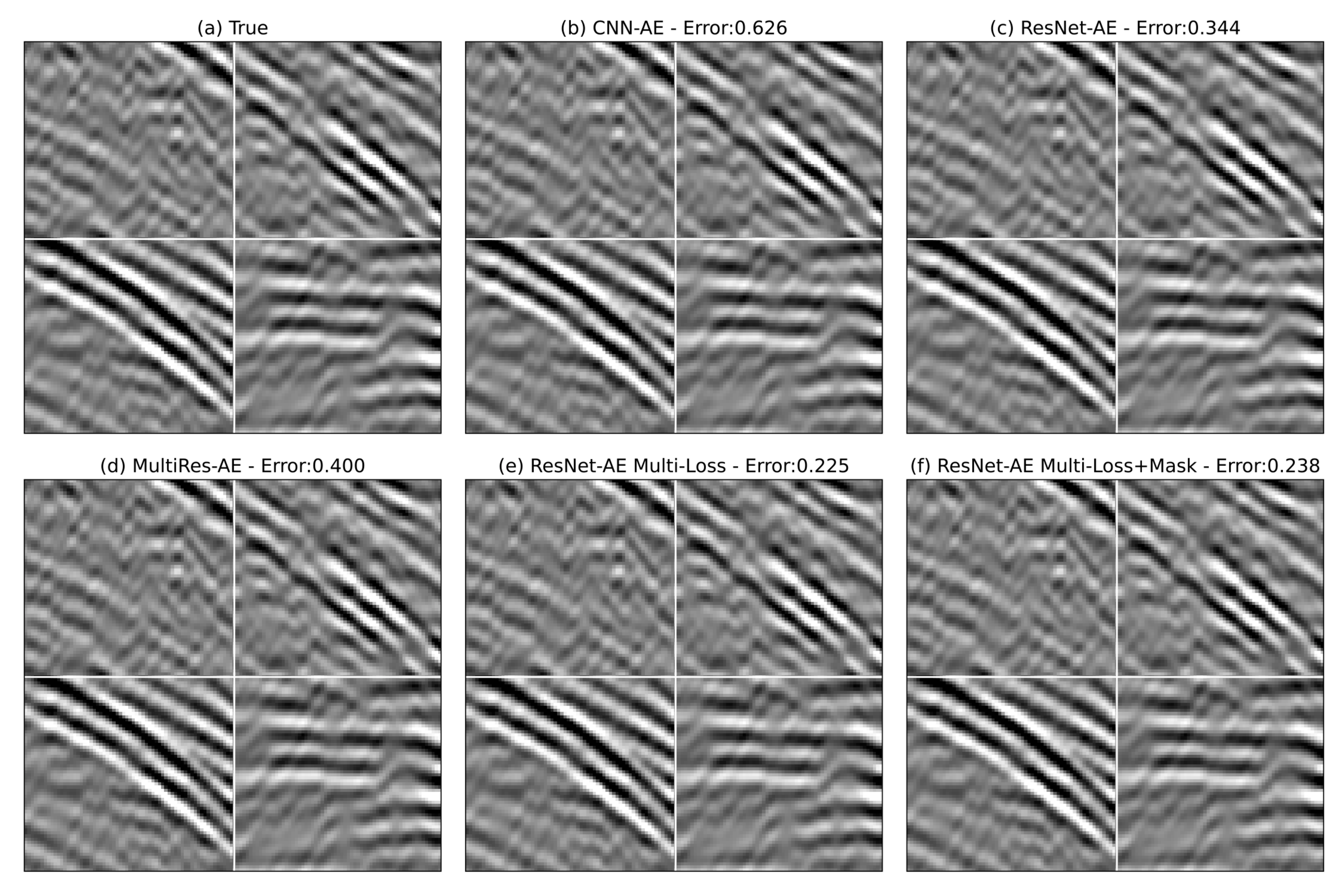}
  \caption{AutoEncoder reconstruction of 4 samples from the validation dataset and MSE computed over the entire validation dataset. a) True, b) Pure convolutional AE with MSE loss and no-preprocessing, c) ResNet AE with MSE loss and no-preprocessing, d) MultiRed AE with MSE loss and no-preprocessing, d) ResNet AE with Multi-task loss and no-preprocessing, and e) ResNet AE with Multi-task loss and masking of input samples.}
  \label{fig:areec}
\end{figure*}

\begin{figure*}
  \centering
  \includegraphics[width=0.99\textwidth]{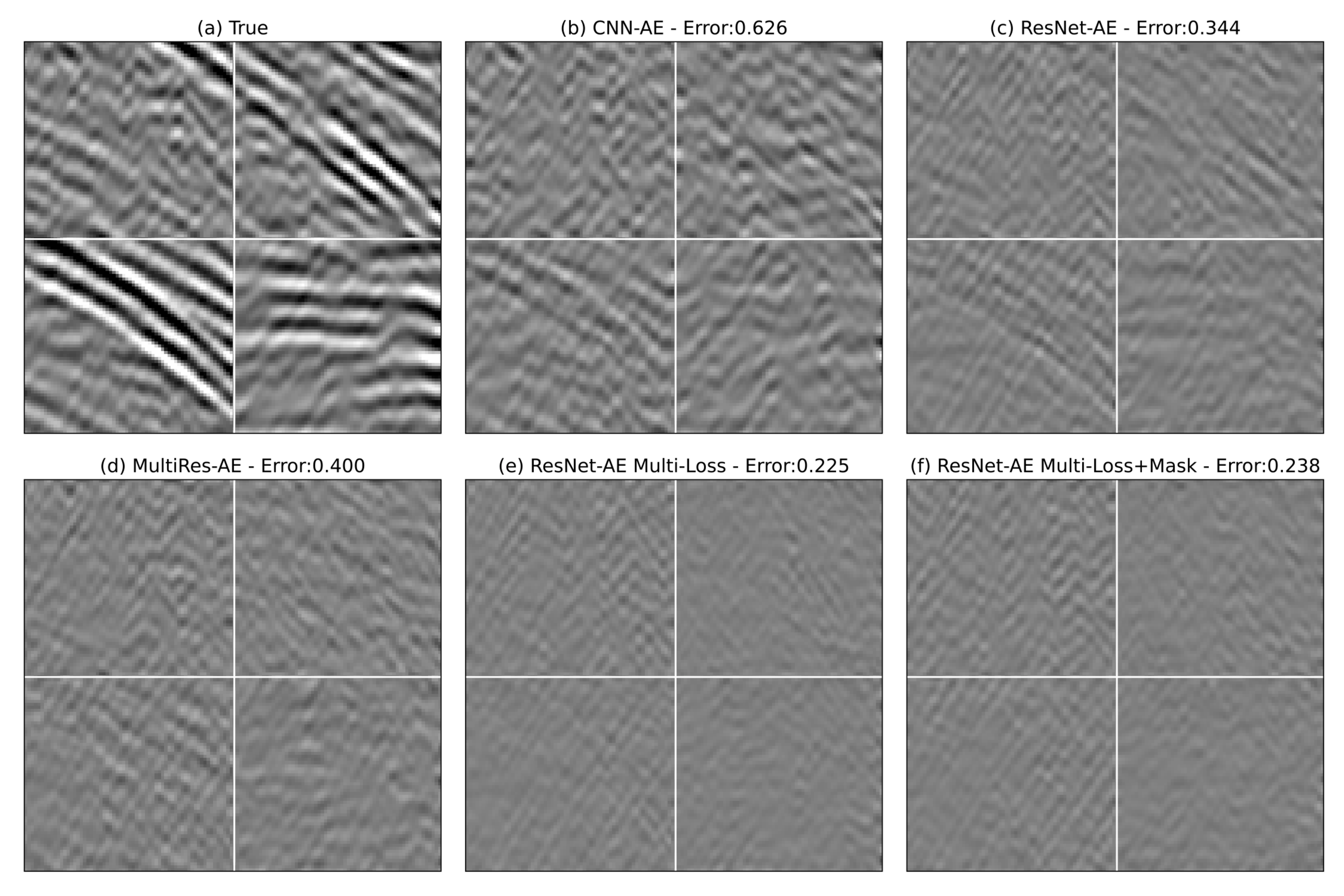}
  \caption{AutoEncoder reconstruction error of 4 samples from the validation dataset and MSE computed over the entire validation dataset. Panels are organized as Figure \ref{fig:areec}.}
  \label{fig:aeerror}
\end{figure*}

Figures \ref{fig:areec} and \ref{fig:aeerror} display four randomly selected patches from the validation dataset of the true and predicted data for the different training strategies, as discussed in the previous section. Moreover, the MSE over the entire validation dataset is reported in each subplot title. From these results, we can clearly observe an improvement in terms of the overall reconstruction capabilities when moving from a purely convolutional AE with single loss to an AE with ResNet blocks and multiple losses. On the one hand, a more sophisticated network architecture with MultiRes blocks led to poorer reconstruction and it is therefore dropped from subsequent analyses. Similarly, a slight decrease in the overall reconstruction error is observed when introducing the masking procedure in the pre-processing of each patch. However, as we will see later, the masking approach seems to help in producing stronger latent codes when it comes to the ultimate goal of using the trained decoder as preconditioners in the solution of different seismic data processing tasks.

\begin{figure*}
  \centering
  \includegraphics[width=0.99\textwidth]{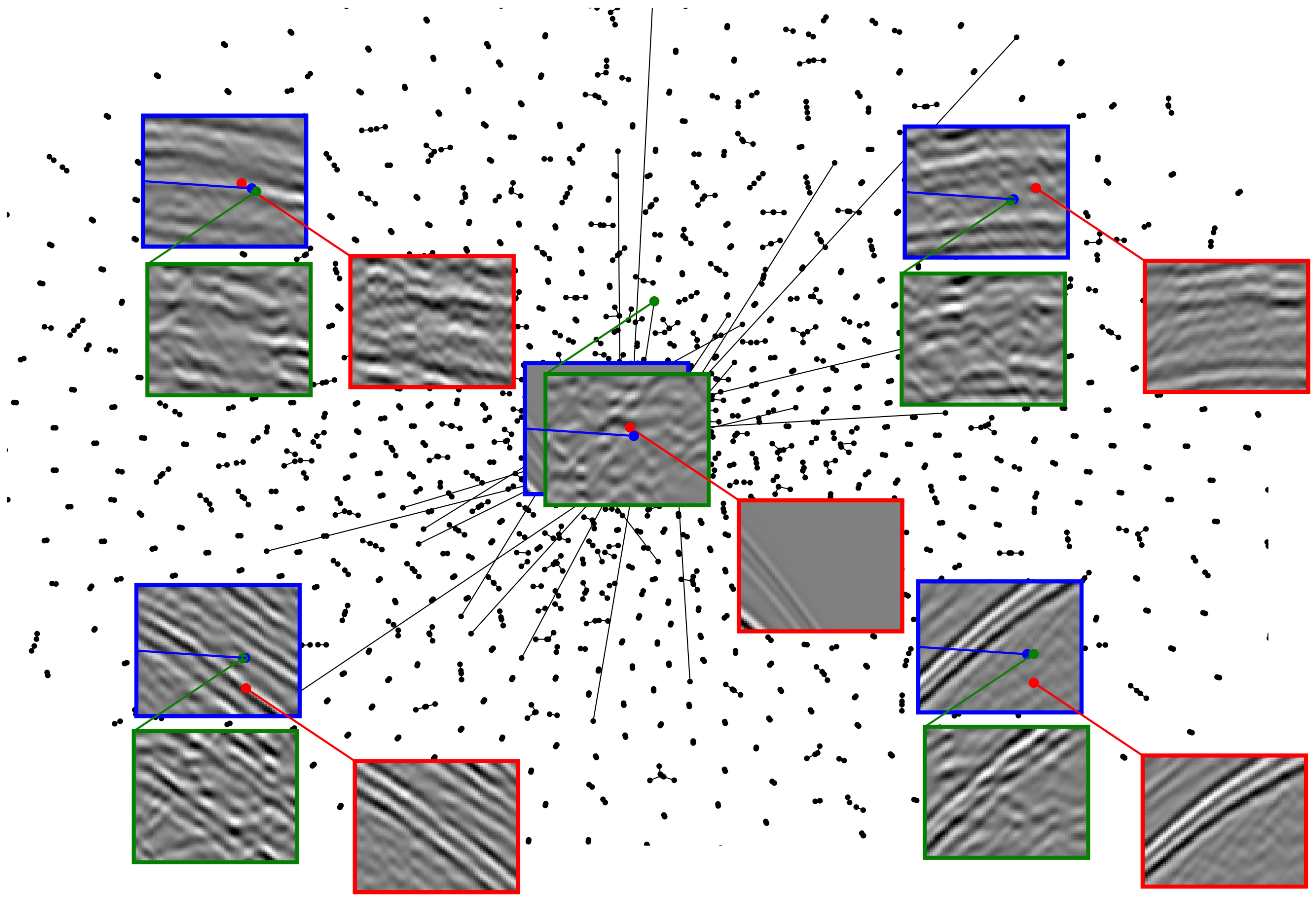}
  \caption{t-SNE visualization of the validation dataset in Figure \ref{fig:areec}f in a two-dimensional space. Seismic patches in blue, red, and green squares correspond to a number of selected validation samples, their closest neighbour and a perturbed version of it in the AE latent space, respectively.}
  \label{fig:tsne}
\end{figure*}

Finally, for the network in Figure \ref{fig:areec}f the latent representations of the entire validation dataset are further compressed to a bi-dimensional space using the t-SNE algorithm \cite{Roweis2002}. This allows us to display them in a scatter plot as shown in Figure \ref{fig:tsne} and analyse how patches of the training seismic data with different features distribute in the latent space. Five points are selected in different areas and their associated  patches are displayed inside blue squares. Similarly, patches in the validation data associated with the closest point in the bi-dimensional space are also displayed inside red squares. We can clearly observe how patches with similar features (e.g., high-amplitude hyperbolic events) cluster together. Finally a small random perturbation is added to the latent representations associated with the five selected patches and the resulting $\textbf{z}$ vectors are fed into the trained decoder. The predicted patches are displayed inside the green squares. We can clearly observe that even areas of the latent space that have not been explored during training lead to representative seismic-looking patches. This is an important result as the subsequent inversion process will operate directly in the latent space and no constraint will be added to enforce the final latent vectors to be in any of the previously sampled positions of such a manifold.

\subsection{Deghosting and interpolation of synthetic dataset}
Joint receiver-side deghosting and data reconstruction is applied here to the first synthetic seismic dataset. The deghosting process can be described as an inverse problem by defining the following modelling operator (e.g., \cite{Grion2017}):
\begin{equation}
\label{eq:deghosting}
\textbf{p}_{-d}=(\textbf{I}+\boldsymbol\Phi)\textbf{p}^-,
\end{equation}
where the model vector $\textbf{p}^-$ contains by the up-going component of the recorded seismic data, whilst the data vector $\textbf{p}_{-d}$ is represented the total pressure wavefield deprived of its direct wave. As far as the physical modelling operator is concerned, $\boldsymbol\Phi$ represents a frequency-wavenumber phase shift operator and $\textbf{I}$ is the identity operator. Combining this modelling operator with the definition of deep preconditioned inversion in equation \ref{eq:aeprecinverse}, we obtain:

\begin{equation}
\label{eq:deghostinginverse}
\hat{\mathbf{z}} = \underset{\mathbf{z}} {\mathrm{argmin}} ||\textbf{p}_{-d} - \mathbf{R} (\textbf{I}+\boldsymbol\Phi) \textbf{W} \textbf{S}  D_\theta( \mathbf{z})||_2^2 + \epsilon_P ||\mathbf{z}||_2^2.
\end{equation}
Since training is performed on patches of size $64 \times 64$, $\textbf{z}$ is composed of a stack of multiple latent space vectors that are decoded by the decoder $D_\theta$, re-scaled from the dynamical range of $(-1, 1)$ used in the network to the actual range of the seismic data via the operator $\textbf{S}$, and finally assembled together by means of a patching operator $\textbf{W}$ (Figure \ref{fig:workflow}b). The scaling factors applied to each patch by the operator $\textbf{S}$ are computed upfront from the corresponding patches in the data $\textbf{p}_{-d}$: whilst these values may not correspond exactly to those of the sought solution, this choice revealed to be robust in all of the scenarios presented in this paper. Finally, the starting guess for the L-BFGS solver is here chosen as follows: $\textbf{z}_{0}=E_\phi(\textbf{S}^{-1}\mathbf{W}^H\mathbf{R}^H\textbf{p}_{-d})$. In other words, the recorded data is divided into patches, each patch is scaled to the dynamic range expected by the network by $\textbf{S}^{-1}$, and then fed into the encoder.

\begin{figure*}[!ht]
  \centering
  \includegraphics[width=0.9\textwidth]{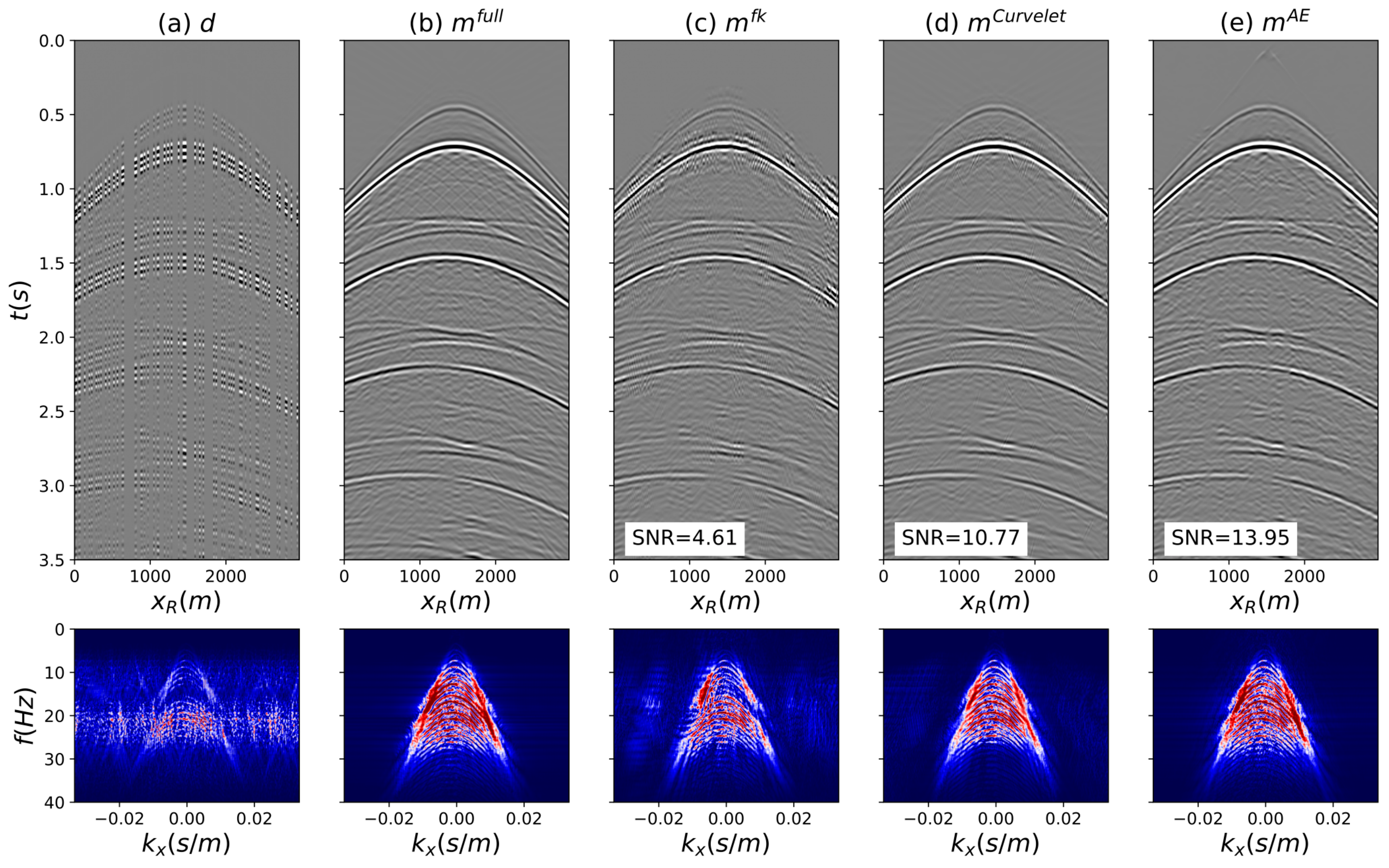}
  \caption{Joint deghosting and reconstruction for irregularly sampled data. a) Subsampled data, b) Benchmark deghosted data, c-d-e) Deghosted and recostructed data using F-K, Curvelet, and AE preconditioners, respectively. All data are shown in time-space domain in the top row and frequency-wavenumber domain in the bottom row.}
  \label{fig:irreg1}
\end{figure*}

\begin{figure*}[!ht]
  \centering
  \includegraphics[width=0.9\textwidth]{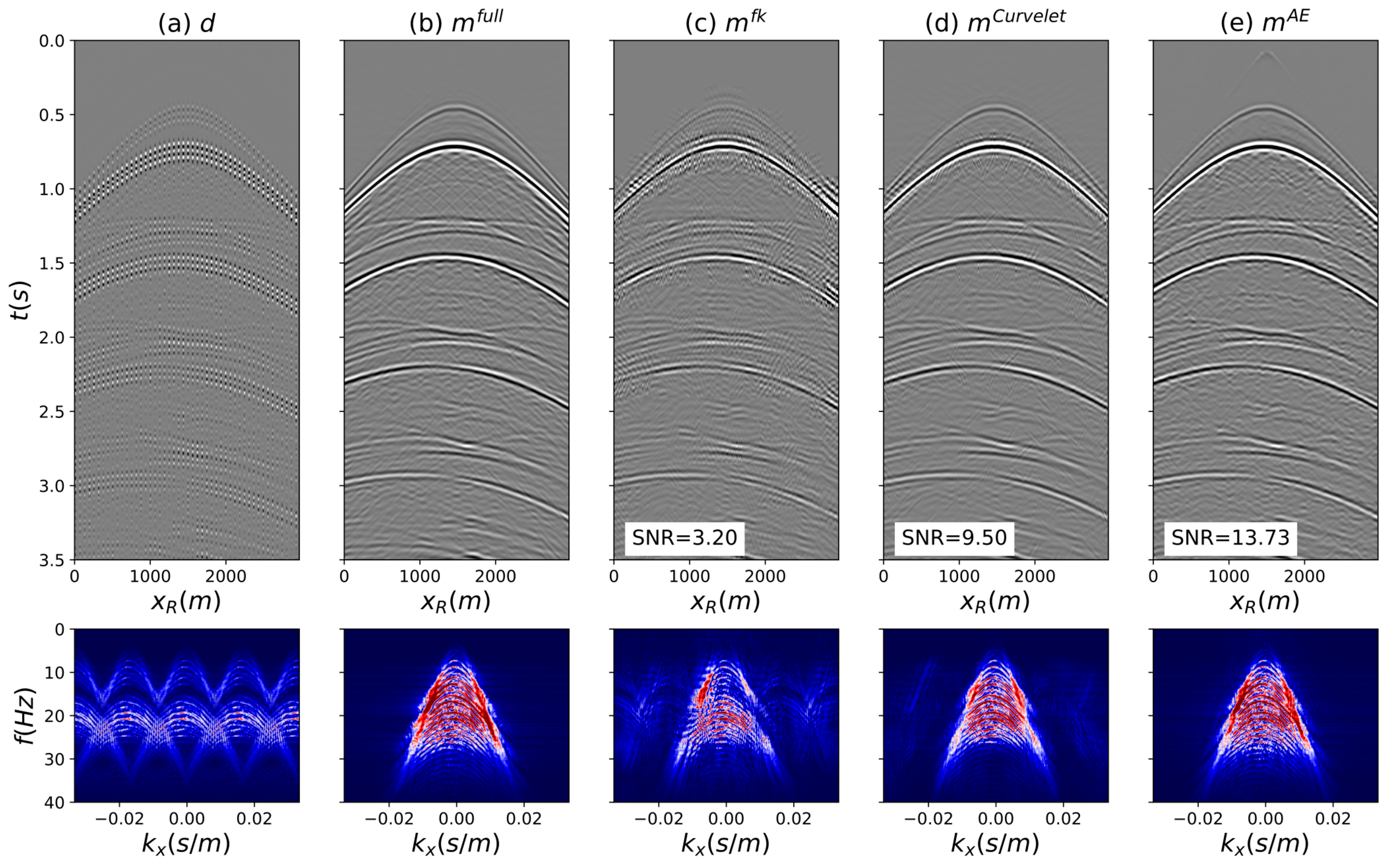}
  \caption{Joint deghosting and reconstruction from regularly sampled data. Panels are in the same order as those in Figure \ref{fig:reg2}.}
  \label{fig:reg1}
\end{figure*}

\begin{figure*}[!ht]
  \centering
  \includegraphics[width=0.99\textwidth]{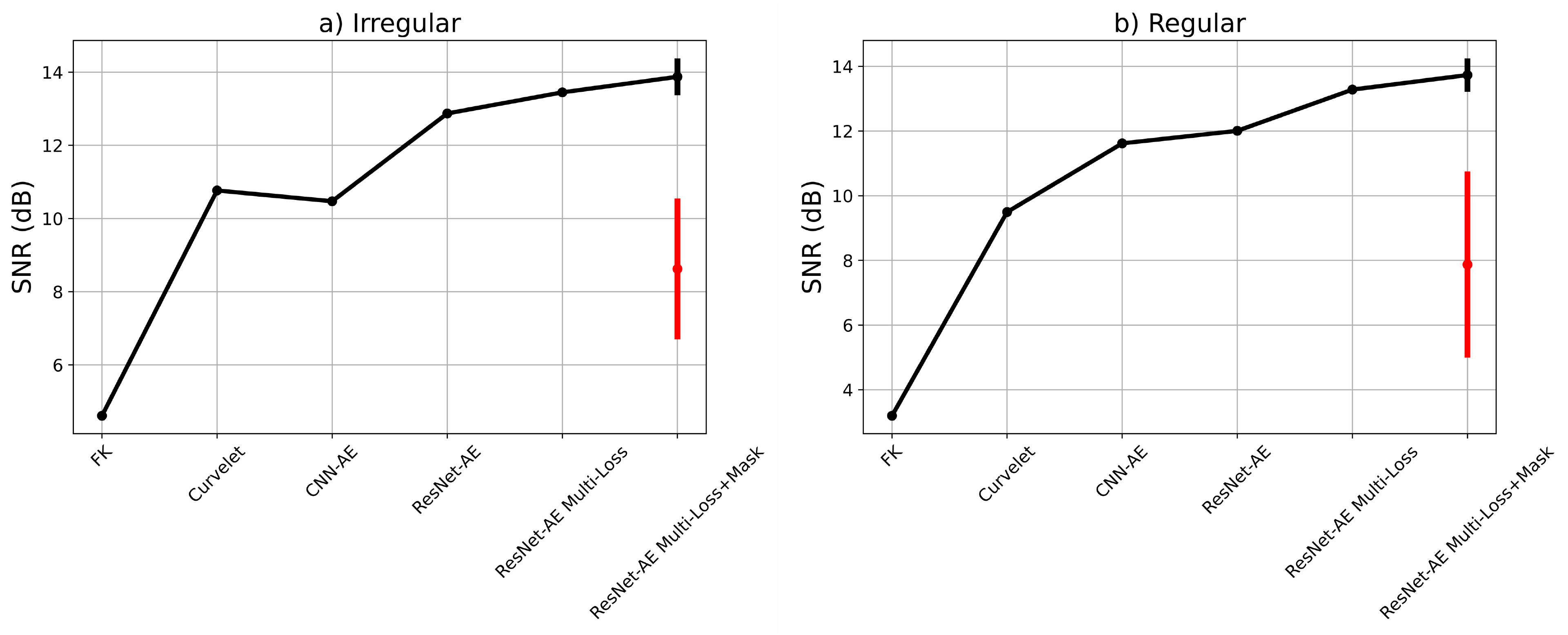}
  \caption{Signal-to-noise ratio as function of different inversion algorithms for the irregular (left) and regular (right) subsampling scenarios. Vertical black lines refer to the standard deviation of 5 inversion using networks with different weight initialization. Vertical red lines refer to the standard deviation of 5 inversion using the same networks and randomly initialized latent codes.}
  \label{fig:snrs}
\end{figure*}

%In this numerical example, receivers are subsampled in two different ways: i) irregularly by a factor of 30\%, ii) regularly by keeping one receiver every 4 (25\% available data). 
Deghosting is initially applied to the fully sampled data for a source in the middle of the array (Figures \ref{fig:irreg1}b and \ref{fig:reg1}b). The ill-posed nature of the problem, due to notches in the F-K spectrum of the data, is mitigated by using a preconditioned inversion with a Curvelet sparsifying transform. The FISTA solver \cite{Beck2009} is used to optimize the associated functional for a total number of 200 iterations. This ensures that we accurately deghost the data also in areas with small amplitude events such that we can use this estimate as our benchmark solution. The subsampled data in Figure \ref{fig:irreg1}a and \ref{fig:reg1}a are inverted with fixed-basis sparsifying transforms, namely the F-K transform in overlapping time-space patches (Figures \ref{fig:irreg1}c and \ref{fig:reg1}c) and the Curvelet transform (Figures \ref{fig:irreg1}d and \ref{fig:reg1}d). Once again we use the FISTA solver for 80 iterations. Finally the trained decoder is used as preconditioner in equation \ref{eq:aeprecinverse}, which is minimized with 80 iterations of L-BFGS (Figure \ref{fig:irreg1}e and \ref{fig:reg1}e). Assuming that the cost of the linear transforms is similar to that of the decoder and having set the number of iterations to 80 for both inversions, allows us to also compare the converge properties of the different algorithms. Both visually and by means of the SNR metric, we conclude that the AE-based inversion produces results of higher quality compared to those from commonly used fixed-basis transforms. Moreover, Figure \ref{fig:snrs} displays the SNRs of the reconstructed upgoing wavefields for the different training strategies discussed in the previous section. We observe how in both cases (irregular and regular), the vanilla CNN AE produces reconstructions on par with those from the sparse inversion with Curvelet transform. However, when the different improvements in the AE training process are introduced the overall SNRs improve by 2-3dB compared to the initial scenario. Moreover, to verify that the initialization of the weights and biases of the network does not have a major impact in the overall quality of reconstruction of the downstream processing task, 5 networks are trained with different initialization and their different decoders are used to perform deghosting and interpolation. The standard deviation of the corresponding SNRs is displayed as a vertical black bar in Figure \ref{fig:snrs}. Finally, to validate the importance of choosing a representative starting guess $\textbf{z}_{0}$, the same deghosting and interpolation process is performed using randomly initialized vectors  $\textbf{z}_{0}$ (within the extected dynamic range of the latent code). The corresponding average and standard deviation SNRs are displayed as red bars in Figure \ref{fig:snrs}, showing a clear decrease in performance likely due to the nonlinear nature of the inverse process and the fact that the starting guess is further away from the optimal solution compared to that using the proposed initialization strategy.

\begin{figure*}
  \centering
  \includegraphics[width=0.99\textwidth]{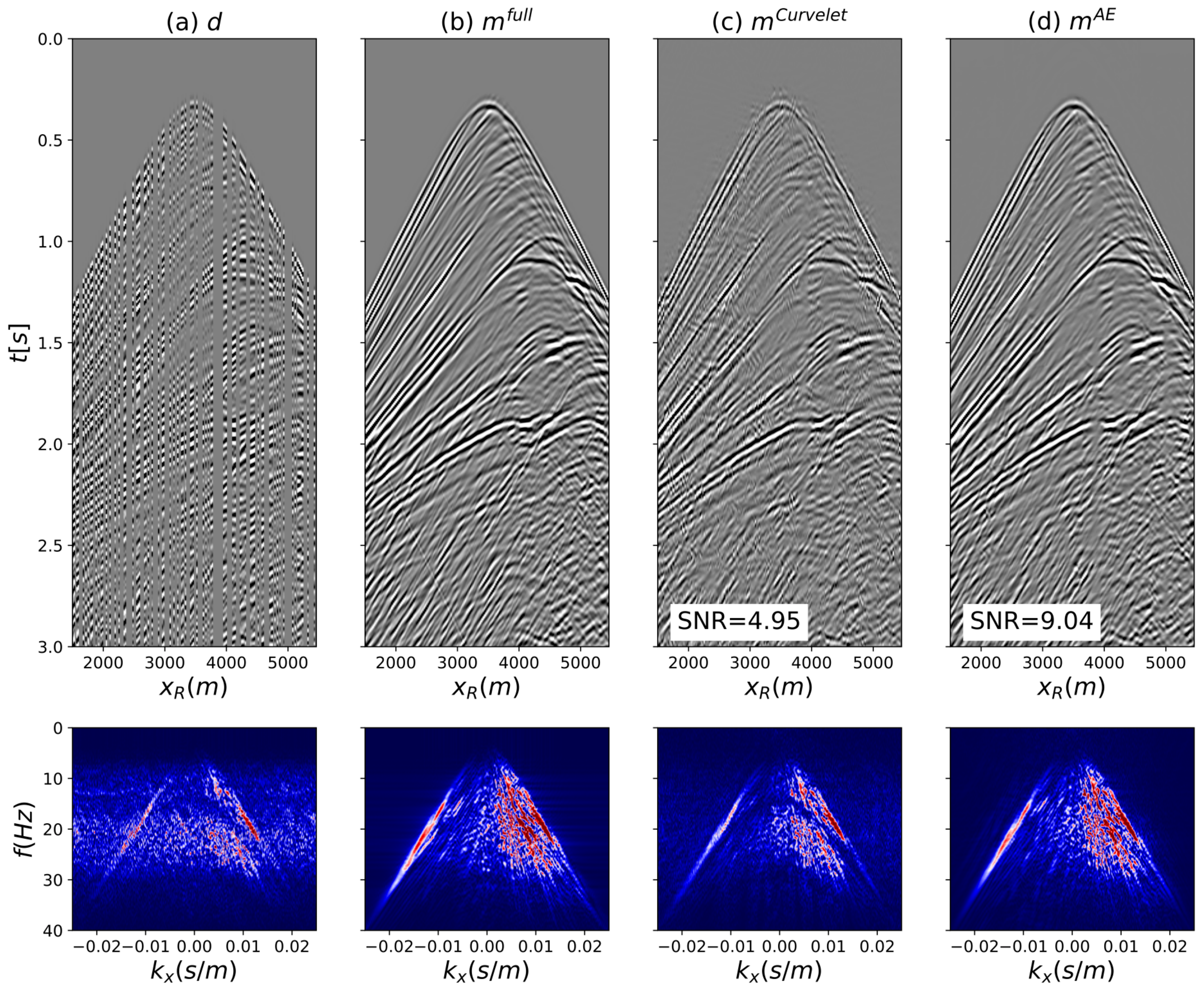}
  \caption{Joint deghosting and reconstruction for irregularly sampled data from the Marmousi model. a) Subsampled data, b) Benchmark deghosted data, c-d-e) Deghosted and recostructed data using F-K, Curvelet, and AE preconditioners, respectively. All data are shown in time-space domain in the top row and frequency-wavenumber domain in the bottom row.}
  \label{fig:irreg2}
\end{figure*}

\begin{figure*}
  \centering
  \includegraphics[width=0.99\textwidth]{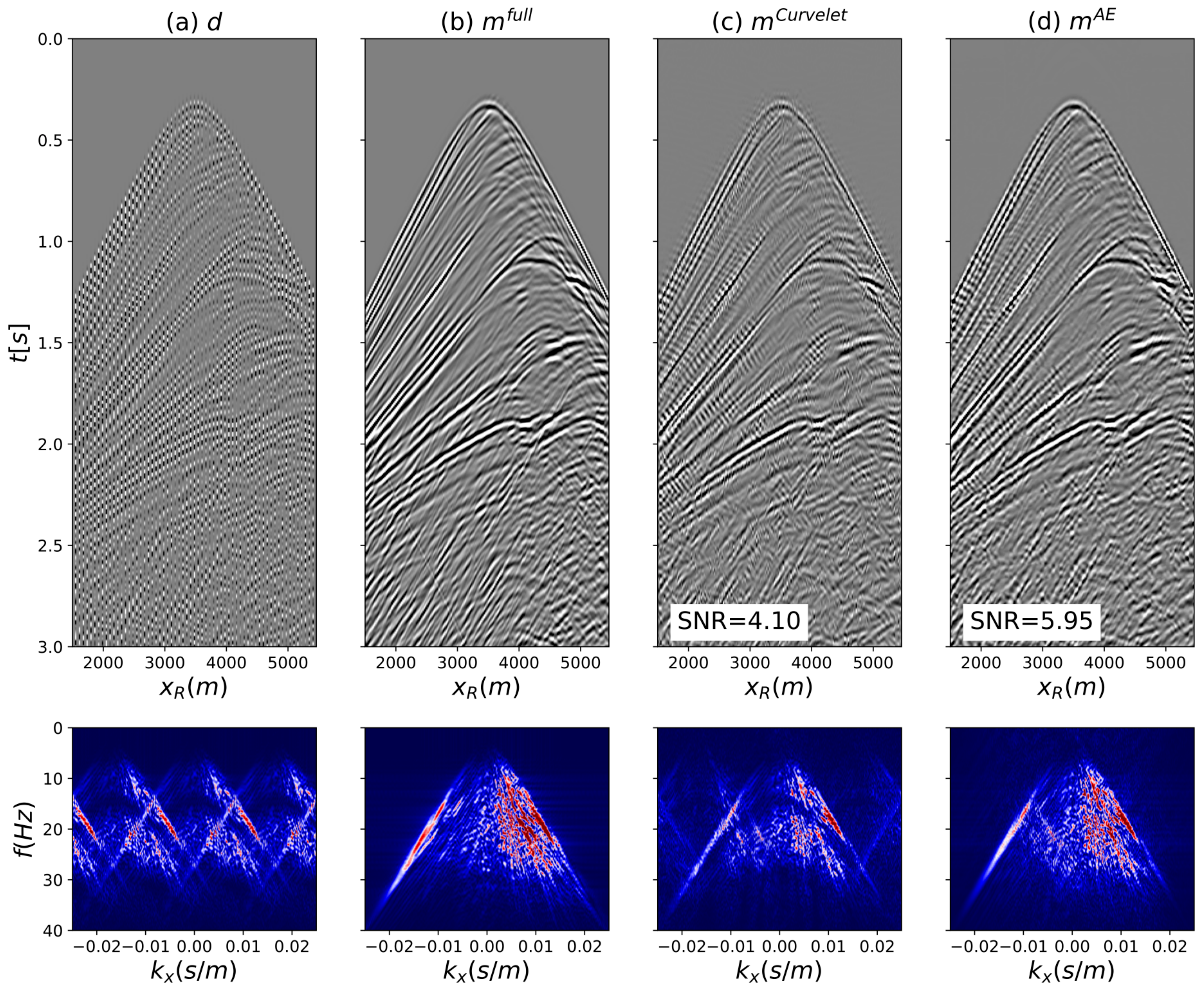}
  \caption{Joint deghosting and reconstruction for regularly sampled data from the Marmousi model. Panels are in the same order as those in Figure \ref{fig:irreg2}.}
  \label{fig:reg2}
\end{figure*}

We move now onto the second synthetic dataset. The training process is performed by creating patches of size $64 \times 64$ in the common receiver domain using the strategy that led to the best results in the previous example. Despite the complexity of the recorded wavefield, the AE network trained with a latent code of size $z=300$ is able to capture a strong representation of the seismic data, producing a decoder with strong interpolation capabilities. Figures \ref{fig:irreg2} and \ref{fig:reg2} display the deghosted and reconstructed upgoing wavefields (panels d) alongside with the subsampled data (panels b), the benchmark deghosting with fully sampled data (panels c), and the reconstruction by means of sparsity-promoting inversion with Curvelet transform. In this example, we have dropped the patched Fourier transform since it proved to be subpar compared to the other two cases.[ht]

\subsection{Wavefield separation and interpolation of synthetic dataset}
We extend the methodology presented in the previous section to the problem of joint reconstruction and wavefield separation; this differs from the former application in that we seek to find both the up- and down-going pressure wavefields that explain the recorded multi-component seismic data. Formally, wavefield separation can be cast as an inverse problem as follows \cite{Wapenaar1998, Neut2012}:

\begin{equation}
\label{eq:wavsep}
\left[\begin{array}{c}
\mathbf{p} \\
\mathbf{v}_{\mathbf{z}}
\end{array}\right]=\left[\begin{array}{cc}
\mathbf{I} & \mathbf{I} \\
\mathbf{W}^{+} & \mathbf{W}^{-}
\end{array}\right]\left[\begin{array}{l}
\mathbf{p}^{+} \\
\mathbf{p}^{-}
\end{array}\right] \rightarrow \textbf{d}=\textbf{Gp}^\pm,
\end{equation}
where $\textbf{p}$ and $\mathbf{v}_{\mathbf{z}}$ are the recorded pressure and vertical particle velocity data, $\textbf{p}^-$ and $\mathbf{p}^+$ are the up- and down-going separated data, $\textbf{I}$ is the identity operator, and $\mathbf{W}^{\pm}=\mathbf{F}^{\mathbf{H}} \operatorname{diag}\left\{\pm \mathbf{k}_{\mathbf{z}} / \rho \boldsymbol\omega \right\} \mathbf{F}$ are operators that perform two~-~dimensional Fourier transforms ($\textbf{F}$) followed by scaling with the obliquity factor in the frequency-wavenumber domain and inverse Fourier transform ($\textbf{F}^H$). A deep preconditioned solution to the above equations can be written as:
\begin{equation}
\label{eq:aewavsep}
\hat{\mathbf{z}}^+, \hat{\mathbf{z}}^- = \underset{\mathbf{z}^+, \mathbf{z}^-} {\mathrm{argmin}} \left\| \mathbf{d} - \mathbf{G} \left[ \begin{array}{l} 
D_\theta(\mathbf{z}^+) \\
D_\theta(\mathbf{z}^-)
\end{array} \right] \right\|_2^2 + \epsilon_Z (||\mathbf{z}^+||^2_2 + ||\mathbf{z}^-||^2_2),
\end{equation}
where the latent vectors of the up- ($\textbf{z}^-$) and down-going ($\textbf{z}^+$) wavefields are simultaneously estimated.

\begin{figure*}
  \centering
  \includegraphics[width=0.99\textwidth]{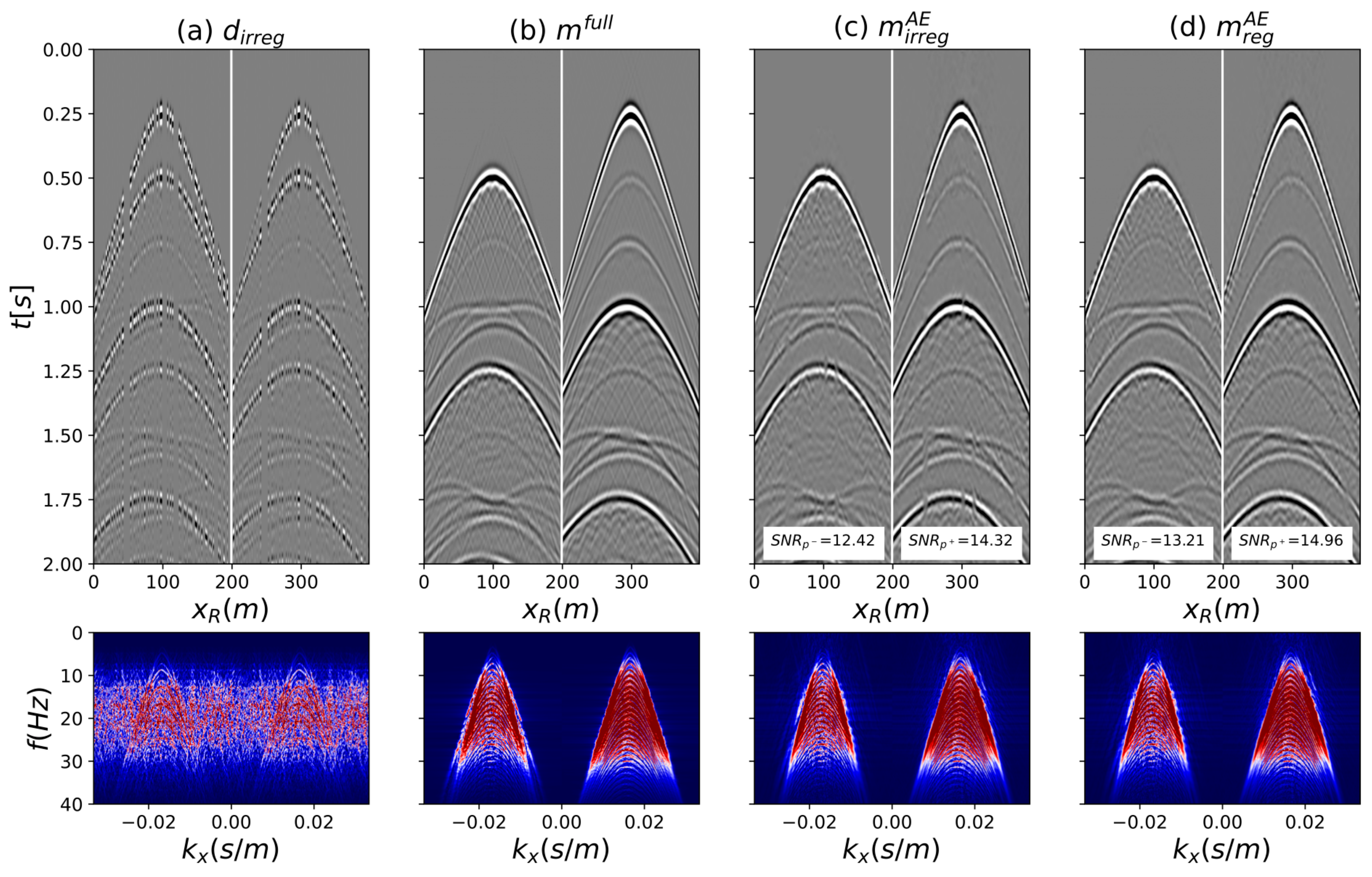}
  \caption{Joint wavefield separation and reconstruction for the synthetic data in ocean-bottom cable configuration. a) Irregularly subsampled data, b) Benchmark wavefield separated data, c-d) Wavefield separated and reconstructed using the AE preconditioner for irregularly and regularly sampled data, respectively. All data are shown in time-space domain in the top row and frequency-wavenumber domain in the bottom row. Pressure and vertical particle velocity components are juxtaposed in panel a, whilst up- and down-going separated components are juxtaposed in all other panels.}
  \label{fig:irreg3}
\end{figure*}

The proposed approach is tested on the first synthetic dataset using multi-component receivers along the seafloor and the same two subsampling strategies. In the training phase, the pressure recordings are sorted in the common receiver gather (CRG) and patches of size $64 \times 64$ are fed to the AE network with ResBlock and multiple losses using the same training strategies as in the deghosting example. The trained decoder is finally combined with the physical modelling operator to reconstruct the missing receivers and separate the up- and down-going components of the data: the estimated wavefields for irregular and regular subsampling are shown in Figure \ref{fig:irreg3}c and d, respectively. In both cases, the reconstructed wavefields closely resemble those obtained by performing a standard wavefield separation on the original, finely sampled data (Figure \ref{fig:irreg3}b). Once again, we observe that selecting the vectors $\textbf{z}^+$ and $\textbf{z}^-$ by feeding a crude estimate of $\textbf{p}^+$ and $\textbf{p}^-$ obtained via simple summation (or subtraction) of the multi-component data to the encoder leads to much better reconstruction compared to using a random or zero starting guess.

\subsection{Deghosting and interpolation of field dataset}
Finally, we consider the Mobil AVO field dataset. The dataset is resorted in the common receiver domain (by extracting all pairs of traces for all sources corresponding to receivers at fixed geographical locations) and divided into 52000 patches of size $64 \times 64$. Note that due to some irregularities in the source array, some of the patches present a small number of missing traces: no attempt is made to fill in such traces prior to training. 

\begin{figure*}
  \centering
  \includegraphics[width=0.99\textwidth]{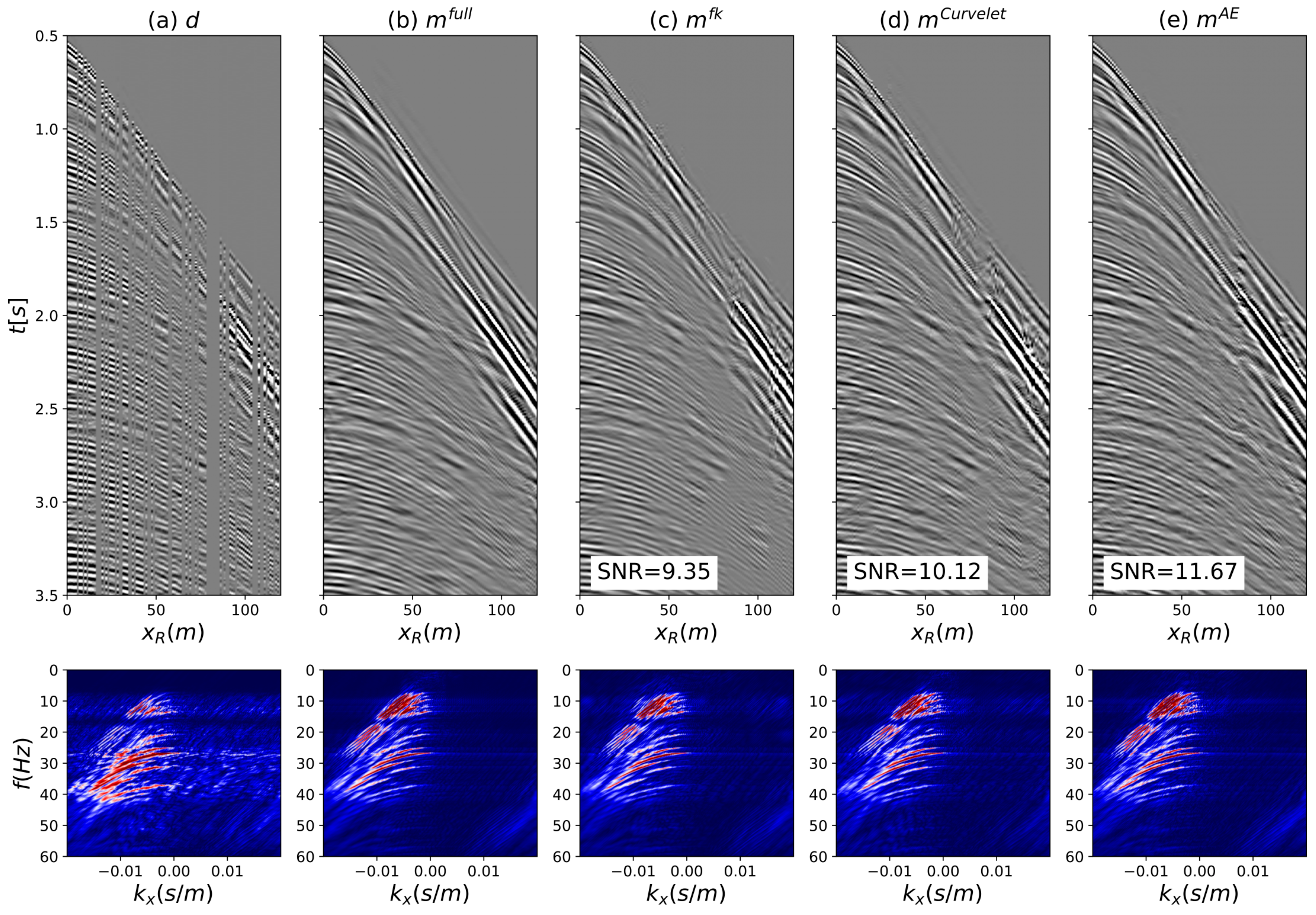}
  \caption{Joint deghosting and reconstruction for the field dataset with irregularly subsampled data. a) Subsampled data, b) Benchmark deghosted data, c-d-e) Deghosted and reconstructed using F-K, Curvelet, and AE preconditioners, respectively. All data are shown in time-space domain in the top row and frequency-wavenumber domain in the bottom row.}
  \label{fig:irreg4}
\end{figure*}

\begin{figure*}
  \centering
  \includegraphics[width=0.99\textwidth]{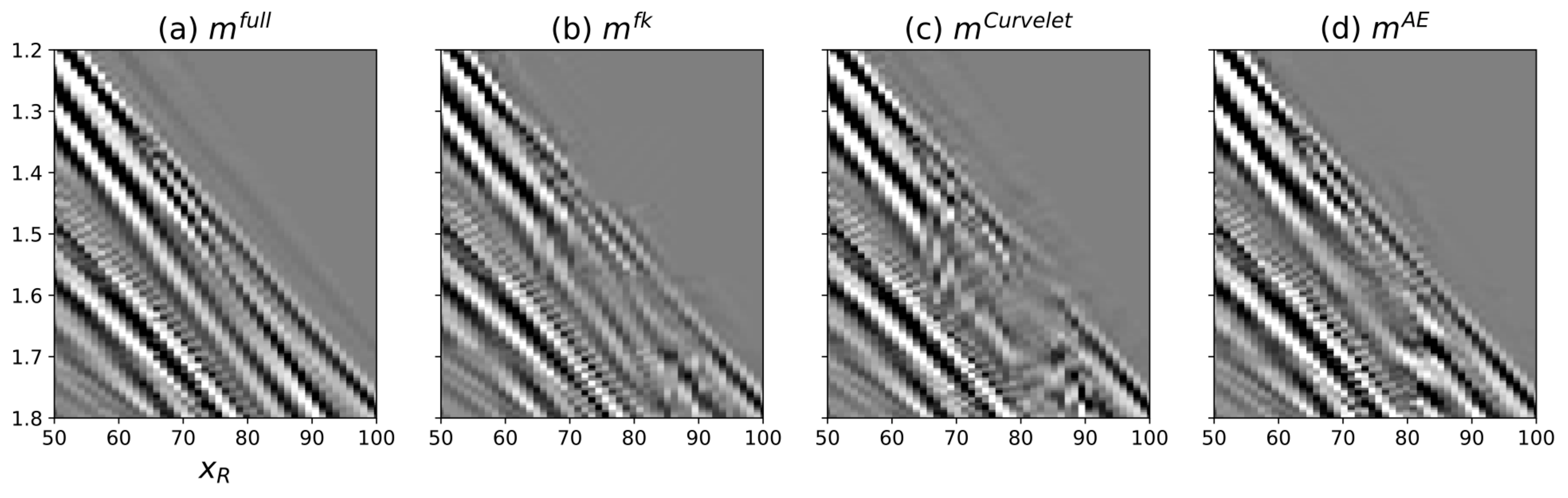}
  \caption{Close-ups of Figure \ref{fig:irreg4} in an area around the direct arrival in the presence of a large gap.}
  \label{fig:irreg4zoom}
\end{figure*}

The training process is carried out using the network, loss, and pre-processing strategy that performed best for the synthetic examples. Receivers are then subsampled irregularly by retaining 60\% of the original array and joint deghosting and interpolation is performed on a randomly selected shot gather. Figure \ref{fig:irreg4} displays the reconstructed upgoing wavefield for the sparsity-promoting inversion with patched Fourier (panel c) and Curvelet (panel d) transforms as well as the reconstruction using the trained deep preconditioner (panel e). When compared to the benchmark solution, we can clearly observe that the inverted wavefield aided by deep preconditoners has more naturally looking seismic events and fewer artefacts than those from the fixed-basis counterparts. Moreover, the deep preconditioned inversion provides an improved reconstruction of the direct arrival (see also close-up in Figure \ref{fig:irreg4zoom}): this result remarks once again the importance of learned transforms that can capture important features from the dataset at hand.

\section{Discussion}
Deep Preconditioners represent an effective tool to regularize severely ill-posed inverse problems like those usually encountered in geophysical data processing. As shown in the Numerical examples section, by learning the characteristic features of seismic data, an AE network provides a nonlinear transformation (i.e., its decoder) that is suitable to aid the solution of seismic processing tasks such as interpolation or wavefield separation. This is further shown to outperform sparsity-promoting inversion with state-of-the-art fixed, linear bases, such as those obtained from the F-K or Curvelet transforms. 

The success of the proposed approach does however rely on the availability of suitable training data and the definition of an appropriate training pipeline. Different strategies could be adopted to define the training dataset. In this work, we have shown how the dataset that we wish to process could be sorted into a domain whose seismic features are similar to those expected in the outcome of the processing step of interest. For example, when seismic data are acquired using ocean-bottom-cable acquisition systems, sources are usually well sampled in space whilst receivers are usually deployed further apart (at least in the crossline direction). By leveraging reciprocity, data sorted along receivers (i.e., common receiver domain) can be used to learn a strong representation of seismic data that are finely sampled in the spatial direction; the learned decoder can be subsequently deployed to reconstruct the missing receivers into a regular and finely sampled grid equal to that of the available sources. Note that in our synthetic example of joint interpolation and wavefield separation, despite the fact the input data used to train the AE network has not been previously decomposed into its up- and down-going constituents, the features learned from such data are shown to be representative of the output data (i.e., separated wavefields) and therefore the decoder of the network can be successfully used as preconditioner to the interpolation and decomposition process. Alternatively, a dataset acquired in a nearby field or during a previous acquisition campaign may be used as input to the AE network. In this case, the chosen dataset must have a more favourable acquisition design, i.e. sources and/or receivers are acquired over a finer spatial grid. In the field data example, to mimic such a scenario we have divided the recorded dataset into two subsets and used the first to train the AE network with the aim of recovering missing receivers in the latter subset. When dealing with streamer data, this is the only viable strategy when we wish to regularize the data along the receiver coordinate: in fact, when receivers are randomly missing or sampled along a coarse regular grid, data sorted in the common receiver domain will also be missing some traces associated with source-receiver pairs that are not sampled, due the fact that receivers move alongside with sources. On the other hand, we note that if our interest is that of recovering missing sources (whilst having access to a finely sampled receiver grid), the first strategy can be employed both for the streamer and ocean-bottom-cable scenarios. In this case, resorting the data in the common source domain provides us with regularly sampled data that can be used to train an AE network to learn useful latent representations. The trained decoder can be ultimately employed to deghost and interpolate the seismic data on the source side.

Recently, a different application of our Deep Preconditoners has been proposed by \cite{Xu2022} in the context of seismic data deblending. By leveraging the fact that blended data in the common source domain present similar features to the deblended data (i.e., coherent seismic events), the authors trained an AE network to learn a robust latent representation from such a data. The decoder is then used in the deblending process to \textit{denoise} the blended data in the common receiver domain, where the blending noise appears as burst-like, trace coherent noise. Since the AE has never seen such with such kind of signal during training, the decoder is naturally encouraged to reproduce only the coherent part of the data during the deblending process. This result highlights the versatility of our approach provided that a suitable training domain can be identified from the available data. Similarly, whilst a single processing task has been carried out in all of the presented examples, another appealing property of the proposed approach lies in the fact that a single learned representation could be used for multiple subsequent tasks. For example, the same representation learned from blended common source gathers could be used to deblend and subsequently interpolate data along the source axis. If successful, this idea may provide a data-centric as opposed to task-centric approach to seismic processing with deep learning where the reliance on training is reduced to a limited number of stages in the processing chain.

Finally, other approaches have recently emerged in the machine learning literature in the context of representation learning. Similar to the AE approach used here, all methods share the common factor of being self-supervised, i.e., do not require labels. Contrastive learning \cite{Liu2021} is one such self-supervised learning technique that has been shown to be able to discover general features of a dataset by simply teaching a model to discriminate between similar and dissimilar training samples. In the spirit of avoiding any manual annotation, data augmentation techniques such as cropping or rotation are used to transform a single input into a number of similar inputs. The model is then fed with both similar and dissimilar pairs and trained to learn to produce latent representations that are close to each other for the first kind of pairs and far away for the other set of pairs. Future work will investigate the suitability of contrastive learning in the context of Deep Preconditioners.

\section{Conclusion}
In this work, we have proposed a general framework to aid the solution of geophysical inverse problems by means of nonlinear, learned preconditioners. Operating in a two-steps fashion, a strong latent representation is first learned from the input seismic data in an unsupervised manner using an AutoEncoder network; the learned decoder is subsequently used to drive the solution of the physics-driven inverse problem at hand. The strength of our approach lies in the fact that no training data is required beyond the input data itself: for example, seismic data containing ghost arrivals are shown to contain useful information that can be later applied to obtained their deghosted counterpart. Different choices of network architectures, loss functions, and pre-processing have been investigated and shown to greatly impact the effectiveness of the representation learning process, and ultimately that of the downstream processing task. More specifically in our numerical examples, the combination of ResNet blocks, multiple losses with learned weights, and masked inputs resulted in deghosting (or wavefield separation) and interpolation capabilities that outperform state-of-the-art sparsity based methods with fixed, linear transformations. Moreover, we observe that the data normalization choice is crucial in ensuring a stable training process and the choice of the initialization is fundamental to achieve a stable inversion process. Finally, the proposed framework may have far wider applicability than the examples of joint reconstruction and wavefield decomposition discussed in this work; other seismic processing steps such as elastic wavefield separation, up/down deconvolution, and target-oriented redatuming will be subject of future studies. 

\section*{Acknowledgments}
I thank KAUST for supporting this research. I am also grateful to Claire Birnie (KAUST) for insightful discussions.

\bibliographystyle{unsrt}  
\bibliography{references}

\end{document}